\documentclass[11pt,
 superscriptaddress,
 amsmath,amssymb,
 aps, physrev,
 floatfix,
]{revtex4-2}

\usepackage{amsmath}
\usepackage{amssymb}
\usepackage{amsthm}
\usepackage{bm}
\usepackage[colorlinks=true]{hyperref}
\hypersetup{linkcolor=red, citecolor=blue, urlcolor=red,
  pdftitle={Nonlinear completions of the Ricci coupling in black-hole scalarization},
  pdfauthor={Subrat Parida}}
\usepackage{orcidlink}
\usepackage{graphicx}
\usepackage{tikz}
\usepackage{dcolumn}
\usepackage{xcolor}
\usepackage{booktabs}
\usepackage{array}
\usepackage{pgfplots}\pgfplotsset{compat=1.18}

\newcommand{\GB}{\mathcal{G}}                 
\newcommand{\hM}{\hat{M}}                     
\newcommand{\phih}{\phi_{h}}                  
\newcommand{\btilt}{\beta_{\mathrm{tilt}}}    
\newcommand{\placeholder}[1]{\textit{[#1]}}   

\begin{document}

\title{Nonlinear completions of the Ricci coupling in black-hole scalarization}

\author{Subrat Parida\,\orcidlink{0000-0001-8754-7719}}
 \email{Corresponding Author: paridasubrat553@gmail.com \\ The author is currently looking for a doctoral position.}{}
\affiliation{Independent Researcher, Puri, Odisha, India}


\begin{abstract}
We study spontaneous scalarization of static, spherically symmetric black holes in
scalar--Gauss--Bonnet gravity with a Ricci coupling
$F=1-\beta\phi^{2}/4+\gamma\beta^{2}\phi^{4}+O(\phi^{6})$ that nonlinearly completes the
quadratic one. For arbitrary couplings and potential, we derive the near-horizon regularity
condition and horizon curvature invariants in closed form; scalarized branches terminate
where this condition degenerates, at finite horizon curvature. A weakly nonlinear expansion
fixes the bifurcation direction exactly: for a massless scalar, its governing coefficient is
quadratic in $\beta$, changes sign at $\beta=1.1156$, and depends linearly on a quartic
self-interaction. At equal horizon values of the scalar, completions differ in mass, to
leading order, in proportion to $\gamma$. A radial-perturbation master equation for
arbitrary couplings confirms the exchange of stability at the bifurcation and turning
points. Consequently, the domain of radially stable black holes shrinks as $\gamma$
increases over the range studied; it is smaller for the exponential completion than for the
quadratic truncation.
\end{abstract}

\maketitle

\section{Introduction}
\label{sec:intro}
Spontaneous scalarization is a mechanism by which a scalar field that is trivial in weak-field
configurations acquires a nontrivial profile in sufficiently compact objects. It was first
identified for neutron stars in scalar--tensor theories~\cite{DEF1993,DEF1996}, where a
tachyonic instability of scalar perturbations in the strong-field region drives the star to a
scalarized configuration while leaving weak-field tests of general relativity (GR) essentially
unaffected. The mechanism extends to black holes: a coupling of the scalar field to the
Gauss--Bonnet invariant $\GB$ destabilizes the Schwarzschild solution below a critical mass,
and branches of scalarized black holes bifurcate from
it~\cite{DonevaYazadjiev2018,Silva2018,AntoniouBakopoulosKanti2018}. Related phenomena occur
for charged black holes~\cite{Herdeiro2018charged} and, for the opposite sign of the
Gauss--Bonnet coupling, for rapidly rotating black holes~\cite{Dima2020}; a comprehensive review
is given in Ref.~\cite{Doneva2024RMP}. Scalarization differs qualitatively from the hair of
dilatonic and shift-symmetric theories~\cite{Kanti1996,SotiriouZhou2014}, in which the black
holes of GR are not solutions. Here, by contrast, the Schwarzschild metric remains a solution
for all values of the couplings, and the scalarized black holes form additional branches that
can be dynamically preferred.

In the minimal model the scalar couples to $\GB$ through a quadratic function. The fundamental
scalarized branch of this model is radially unstable~\cite{BlazquezSalcedo2018,Silva2019} and
therefore cannot describe the endpoint of the tachyonic instability of the Schwarzschild
solution. Radial stability can be restored by nonlinear terms in the Gauss--Bonnet
coupling~\cite{DonevaYazadjiev2018,BlazquezSalcedo2018}, by a quartic self-interaction of the
scalar~\cite{Macedo2019}, or by an additional coupling to the Ricci
scalar~\cite{Antoniou2021,Antoniou2022}; the effect of a scalar mass has also been
investigated~\cite{Doneva2019massive}. The Ricci coupling belongs to the class of terms that
can trigger scalarization at quadratic order in the scalar field~\cite{Andreou2019}. Besides
stabilizing the scalarized black holes, it mitigates~\cite{Thaalba2024,Thaalba2025} the loss
of hyperbolicity that affects the evolution problem in scalar--Gauss--Bonnet
gravity~\cite{RipleyPretorius2019a,RipleyPretorius2019b,EastRipley2021}, and it also affects
the scalarization of neutron stars~\cite{Ventagli2021}. In particular, for a sufficiently
strong Ricci coupling the scalarized branch changes the direction in which it leaves the
Schwarzschild family, and the black holes near the threshold become radially
stable~\cite{Antoniou2021,Antoniou2022}.

From the viewpoint of effective field theory, quadratic couplings are only the leading terms of
coupling functions that generically contain higher powers of the scalar. For the Gauss--Bonnet
coupling, the importance of these terms is well established: they control the radial stability
of the fundamental branch~\cite{BlazquezSalcedo2018}, the order of the scalarization
transition~\cite{Herdeiro2026}, and the properties of rotating scalarized black
holes~\cite{Staykov2025}. In most studies of scalarized black holes, however, the Ricci
coupling is
quadratic~\cite{Antoniou2021,Antoniou2022,Thaalba2024,Thaalba2025,Kleihaus2023,Minamitsuji2024};
a non-quadratic, Gaussian Ricci coupling has been employed in an analysis of hyperbolicity and
of the minimum black-hole mass~\cite{Rossi2026}. To our knowledge, the question of how the
nonlinear completion of the Ricci coupling shapes the existence and the stability of scalarized
black holes has not been addressed systematically. We study this question for the family
\begin{equation}
F(\phi)=1-\frac{\beta}{4}\,\phi^{2}+\gamma\,\beta^{2}\phi^{4}+O(\phi^{6})
\label{eq:intro-family}
\end{equation}
of functions multiplying the Ricci scalar, together with a quadratic Gauss--Bonnet coupling and,
where indicated, a scalar mass and a quartic self-interaction. All members of this family share
the linear physics of the quadratic coupling, including the threshold of the tachyonic
instability, and differ only through nonlinear terms, whose leading effect is controlled by the
dimensionless coefficient $\gamma$. The quadratic truncation corresponds to $\gamma=0$ and the
exponential completion $F=e^{-\beta\phi^{2}/4}$ to $\gamma=1/32$. All theories in this family
belong to the Horndeski class~\cite{Horndeski1974,Kobayashi2011}, so their field equations are
of second order.

Two structural facts organize our analysis. First, near the threshold the scalarized branch
emerges from the Schwarzschild family through a bifurcation from a simple
eigenvalue~\cite{CrandallRabinowitz1971}. Whether the branch initially exists on the side of the
threshold where the Schwarzschild solution is stable (a subcritical bifurcation) or unstable (a
supercritical bifurcation) is fixed by a single coefficient, which also determines the radial
stability of the scalarized black holes near the threshold. This coefficient has so far been
inferred numerically from the shape of the branches~\cite{Antoniou2021,Antoniou2022}.
Landau-type descriptions of the scalarization transition have been developed numerically for
black holes without a Ricci coupling~\cite{Herdeiro2026} and for neutron
stars~\cite{Muniz2025,Unluturk2025}, for which the Landau coefficients have recently been
derived from first principles~\cite{Ozinan2026}. Here we compute the coefficient exactly by a
weakly nonlinear expansion. Second, along a branch of equilibria the radial stability can change
only where a perturbation mode crosses zero frequency, which generically happens at turning
points of the mass~\cite{Sorkin1981}; such behavior has been observed for scalarized black holes
with self-interactions~\cite{Macedo2019} and with a Ricci coupling~\cite{Antoniou2022}.
Together, these two facts relate the domains of existence of the scalarized black holes to
their domains of radial stability.

Radial stability is necessary but not sufficient for stability. For the quadratic Ricci
coupling, radially stable scalarized black holes can be unstable against non-radial
perturbations: quadrupolar~\cite{Kleihaus2023} and hexadecapolar~\cite{BlazquezSalcedo2024GRG}
instabilities have been found, and analyses of high multipoles indicate a hierarchy of angular
instabilities~\cite{Minamitsuji2024,BlazquezSalcedo2026hier}. In this work we restrict
attention to static, spherically symmetric black holes and to radial perturbations. All our
statements about stability therefore refer to radial stability, and we return to the
non-radial problem in Sec.~\ref{sec:discussion}.

Our main results are as follows.
\begin{enumerate}
\renewcommand{\labelenumi}{(\roman{enumi})}
\item For arbitrary coupling functions and potential, we derive the near-horizon regularity
condition in closed form, together with closed-form expressions for the curvature invariants at
the horizon; the regularity condition reduces to the special cases known in the
literature~\cite{Antoniou2021,Macedo2019}. All the branches we compute terminate where the
discriminant of this condition vanishes, while the horizon curvature remains finite. As this
point is approached, the characteristic speed of radial perturbations decreases within a thin
layer outside the horizon.
\item A weakly nonlinear expansion determines the direction of the bifurcation exactly. For a
massless scalar, the coefficient of the leading correction to the mass along the branch is a
quadratic polynomial in $\beta$ that changes sign at $\beta=\btilt\simeq1.1156$, and a quartic
self-interaction contributes a term linear in its strength.
\item All completions coincide through third order in the horizon value $\phih$ of the scalar.
At fourth order, the normalized mass of a completion at fixed $\phih$ is shifted relative to the
quadratic truncation by an amount proportional to $\gamma$, with a coefficient that depends only
on $\beta$. We confirm this law with four completions, two of which were introduced as
independent tests.
\item The ordering of the domains of existence of different completions depends on $\beta$.
Combined with the exchange of radial stability at the bifurcation and at turning points, which
we verify using a master equation for radial perturbations derived for arbitrary couplings, it
yields a definite ordering of the domains of radially stable black holes: these domains shrink
monotonically as $\gamma$ increases over the range studied. In particular, the exponential
completion supports a smaller domain of radially stable scalarized black holes than the
quadratic truncation, both in the absence and in the presence of a scalar mass and
self-interaction.
\item These conclusions are independent of the conformal frame.
\end{enumerate}

The paper is organized as follows. In Sec.~\ref{sec:theory} we introduce the theory, the family
of completions, and our conventions, including the Einstein-frame formulation.
Section~\ref{sec:static} presents the static field equations, the general near-horizon
regularity condition, and the horizon curvature invariants. The weakly nonlinear analysis is
developed in Sec.~\ref{sec:wnl}. Section~\ref{sec:existence} describes the domains of
existence, and Sec.~\ref{sec:mechanism} the mechanism that orders them together with the
completion law. Radial stability and hyperbolicity are analyzed in Sec.~\ref{sec:stability}. We
discuss our results and conclude in Sec.~\ref{sec:discussion}.
Technical details are collected in the Appendixes. Throughout, we use units $G=c=1$ and the
metric signature $(-,+,+,+)$.

\section{Theory, Ricci completions, and conventions}
\label{sec:theory}
\subsection{Action and field equations}
\label{subsec:action}

We consider the action
\begin{equation}
S=\frac{1}{16\pi}\int d^{4}x\,\sqrt{-g}\,\Bigl[F(\phi)R-\frac{1}{2}\,g^{\mu\nu}\partial_{\mu}\phi\,
\partial_{\nu}\phi-V(\phi)+f(\phi)\,\GB\Bigr],
\label{eq:action}
\end{equation}
where
\begin{equation}
\GB=R_{\mu\nu\rho\sigma}R^{\mu\nu\rho\sigma}-4R_{\mu\nu}R^{\mu\nu}+R^{2}
\label{eq:GB}
\end{equation}
is the Gauss--Bonnet invariant. Throughout this work the Gauss--Bonnet coupling and the potential
are
\begin{equation}
f(\phi)=\frac{\alpha}{2}\,\phi^{2},\qquad
V(\phi)=\frac{1}{2}\,\mu^{2}\phi^{2}+\frac{1}{2}\,\lambda\,\phi^{4},
\label{eq:f-V}
\end{equation}
with $\alpha>0$, $\mu\geq0$, and $\lambda\geq0$. The Ricci coupling $F(\phi)$ is specified in
Sec.~\ref{subsec:completions}; it satisfies $F(0)=1$, so that general relativity is recovered
for $\phi=0$. We use the metric signature $(-,+,+,+)$ and the curvature conventions
$R^{\rho}{}_{\sigma\mu\nu}=\partial_{\mu}\Gamma^{\rho}_{\nu\sigma}-\partial_{\nu}\Gamma^{\rho}_{\mu\sigma}
+\Gamma^{\rho}_{\mu\lambda}\Gamma^{\lambda}_{\nu\sigma}-\Gamma^{\rho}_{\nu\lambda}\Gamma^{\lambda}_{\mu\sigma}$
and $R_{\sigma\nu}=R^{\mu}{}_{\sigma\mu\nu}$. Antisymmetrization carries unit weight,
$X_{[\mu\nu]}=\frac{1}{2}(X_{\mu\nu}-X_{\nu\mu})$, and we work in units $G=c=1$; Planck's
constant does not appear, since the analysis is entirely classical.

Variation of Eq.~\eqref{eq:action} with respect to $g^{\mu\nu}$ and $\phi$ yields the field
equations $E_{\mu\nu}=0$ and $E_{\phi}=0$, where
$E_{\mu\nu}\equiv(16\pi/\sqrt{-g})\,\delta S/\delta g^{\mu\nu}$ and
$E_{\phi}\equiv(16\pi/\sqrt{-g})\,\delta S/\delta\phi$ are given by
\begin{align}
E_{\mu\nu}={}&F\,G_{\mu\nu}+g_{\mu\nu}\Box F-\nabla_{\mu}\nabla_{\nu}F
-\frac{1}{2}\,\partial_{\mu}\phi\,\partial_{\nu}\phi+\frac{1}{4}\,g_{\mu\nu}(\partial\phi)^{2}
+\frac{1}{2}\,g_{\mu\nu}V+4P_{\mu\rho\nu\sigma}\nabla^{\rho}\nabla^{\sigma}f,
\label{eq:Emn}\\
E_{\phi}={}&\Box\phi-V'(\phi)+F'(\phi)\,R+f'(\phi)\,\GB .
\label{eq:Ephi}
\end{align}
Here a prime denotes differentiation with respect to $\phi$,
$(\partial\phi)^{2}=g^{\mu\nu}\partial_{\mu}\phi\,\partial_{\nu}\phi$, and
\begin{equation}
P_{\mu\rho\nu\sigma}=R_{\mu\rho\nu\sigma}+2g_{\mu[\sigma}R_{\nu]\rho}+2g_{\rho[\nu}R_{\sigma]\mu}
+R\,g_{\mu[\nu}g_{\sigma]\rho}
\label{eq:P}
\end{equation}
is the double dual of the Riemann tensor. It has the algebraic symmetries of the Riemann tensor,
it is divergence free, $\nabla^{\mu}P_{\mu\rho\nu\sigma}=0$, and its trace is
$P^{\mu}{}_{\rho\mu\sigma}=-G_{\rho\sigma}$. Owing to these properties, Eqs.~\eqref{eq:Emn}
and~\eqref{eq:Ephi} are of second order in derivatives, in accordance with the fact that the
theory belongs to the Horndeski class~\cite{Horndeski1974,Kobayashi2011}. Diffeomorphism
invariance of Eq.~\eqref{eq:action} implies the identity
\begin{equation}
\nabla^{\mu}E_{\mu\nu}=-\frac{1}{2}\,E_{\phi}\,\partial_{\nu}\phi ,
\label{eq:noether}
\end{equation}
which holds for arbitrary field configurations, not only for solutions. We have verified
Eq.~\eqref{eq:noether} for the equations exactly as written above on generic test
configurations (Appendix~\ref{app:covariant}). In particular, the metric equations imply the
scalar equation wherever $\partial\phi\neq0$.

\subsection{Nonlinear completions of the Ricci coupling}
\label{subsec:completions}

In terms of the variable $x\equiv\beta\phi^{2}/4$, the family~\eqref{eq:intro-family} reads
\begin{equation}
F(\phi)=1-x+16\gamma\,x^{2}+O(x^{3}).
\label{eq:F-family}
\end{equation}
The dimensionless parameter $\beta$ sets the strength of the Ricci coupling at quadratic order,
and $\gamma$ is the leading nonlinear datum of the completion. We consider the four completions
listed in Table~\ref{tab:completions}: the quadratic truncation (Q), the exponential completion
(E), and two further completions, a rational one (R) and a polynomial one with $\gamma<0$ (M),
which we introduced as independent tests of the completion law of Sec.~\ref{sec:mechanism}.
When a distinction is needed, $F_{X}$ denotes the completion $X\in\{\mathrm{Q},\mathrm{E},
\mathrm{R},\mathrm{M}\}$.

\begin{table}[t]
\caption{Completions of the Ricci coupling considered in this work, with $x=\beta\phi^{2}/4$. The
coefficient $\gamma$ is defined by $F=1-\beta\phi^{2}/4+\gamma\beta^{2}\phi^{4}+O(\phi^{6})$. The
last column gives the zero of $F$, beyond which the effective gravitational coupling changes
sign.}
\label{tab:completions}
\begin{ruledtabular}
\begin{tabular}{llcl}
Completion & $F(\phi)$ & $\gamma$ & Zero of $F$\\
\hline
Quadratic truncation (Q) & $1-x$ & $0$ & $x=1$\\
Exponential (E) & $e^{-x}$ & $1/32$ & none\\
Rational (R) & $(1+x)^{-1}$ & $1/16$ & none\\
Polynomial with $\gamma<0$ (M) & $1-x-x^{2}/2$ & $-1/32$ & $x=\sqrt{3}-1$\\
\end{tabular}
\end{ruledtabular}
\end{table}

The physical meaning of the completion is transparent in the Jordan frame, where $F(\phi)$
determines the effective gravitational coupling $G_{\mathrm{eff}}=G/F(\phi)$ experienced by the
metric. For $\beta>0$ an excited scalar reduces $F$ and therefore strengthens gravity. All
completions agree on this effect at leading order and differ in its nonlinear saturation,
according to $F_{X}-F_{\mathrm{Q}}=\gamma\beta^{2}\phi^{4}+O(\phi^{6})$: for $\gamma>0$ the
effective coupling grows more slowly with the scalar amplitude than in the quadratic truncation,
whereas for $\gamma<0$ it grows faster. The completions Q and M vanish at finite amplitude, where
$G_{\mathrm{eff}}$ diverges. We restrict attention to configurations with $F(\phi)>0$, for which
$G_{\mathrm{eff}}$ is positive and the Einstein-frame formulation of Sec.~\ref{subsec:einstein}
exists. All solutions constructed in this work satisfy this condition: along every branch,
$F(\phi_{h})>0.2$ on the horizon, and for the most strongly coupled solutions we have verified
that $|\phi|$ decreases monotonically outward, so that $F$ attains its minimum on the horizon.

Since $F'(\phi)=-\beta\phi/2+4\gamma\beta^{2}\phi^{3}+O(\phi^{5})$, in an expansion in the
amplitude of the scalar the parameter $\gamma$ first enters the metric equations~\eqref{eq:Emn}
at fourth order. In the scalar equation~\eqref{eq:Ephi} it multiplies the Ricci scalar, which
vanishes on the Schwarzschild background and is itself of second order, so that $\gamma$ enters
only at fifth order. Solutions of different completions with the same couplings therefore
coincide through third order in the amplitude of the scalar, a fact that underlies the analysis
of Secs.~\ref{sec:wnl} and~\ref{sec:mechanism}.

\subsection{The Schwarzschild black hole and the threshold of scalarization}
\label{subsec:threshold}

Because $F(0)=1$ and $F'(0)=f'(0)=V'(0)=0$, the Schwarzschild metric with $\phi=0$ solves
Eqs.~\eqref{eq:Emn} and~\eqref{eq:Ephi} for all values of the couplings. Linearizing
Eq.~\eqref{eq:Ephi} about this solution and using $R=0$ and $\GB=48M^{2}/r^{6}$ gives
\begin{equation}
\Bigl(\Box-\mu^{2}+\frac{48\,\alpha M^{2}}{r^{6}}\Bigr)\delta\phi=0 .
\label{eq:linear}
\end{equation}
The Ricci coupling drops out entirely: neither $\beta$ nor $\gamma$ affects the threshold of the
instability or the zero mode at the threshold. The effective squared mass,
$\mu^{2}-48\alpha M^{2}/r^{6}$, is most negative on the horizon. For sufficiently large
$\alpha/M^{2}$ this tachyonic well supports a growing, spherically symmetric mode, and the
Schwarzschild solution becomes unstable. For a massless scalar, the fundamental (nodeless)
static zero mode appears at
\begin{equation}
\frac{\alpha}{M^{2}}=0.7256128492627,\qquad\text{that is,}\qquad
\hM_{c}\equiv\frac{M}{\sqrt{\alpha}}\bigg|_{c}=1.1739443708,
\label{eq:threshold}
\end{equation}
a value that we have obtained by two independent methods. It corresponds to $\eta/M^{2}=2.9025$ in
the normalization of Ref.~\cite{Silva2018}, in agreement with the value $2.902$ quoted there. A
scalar mass raises the threshold; for instance, $\alpha/M^{2}=0.85927585$ for $\mu M=0.1$. At the
threshold a branch of scalarized black holes bifurcates from the Schwarzschild family. Since the
linear problem is independent of the Ricci coupling, all completions share the same bifurcation
point and the same zero mode, and they differ only in the nonlinear structure of the branch, which
is the subject of Secs.~\ref{sec:wnl}--\ref{sec:stability}.

\subsection{Einstein frame and frame independence}
\label{subsec:einstein}

For $F>0$, the conformal transformation $\tilde g_{\mu\nu}=F(\phi)\,g_{\mu\nu}$ maps the Ricci,
kinetic, and potential terms of Eq.~\eqref{eq:action} into canonical Einstein-frame form,
\begin{equation}
\sqrt{-g}\,\Bigl[FR-\frac{1}{2}(\partial\phi)^{2}-V\Bigr]
=\sqrt{-\tilde g}\,\Bigl[\tilde R-\frac{1}{2}\,K(\phi)(\tilde\partial\phi)^{2}
-\frac{V}{F^{2}}\Bigr]+\text{total derivative},
\label{eq:einstein-frame}
\end{equation}
where $(\tilde\partial\phi)^{2}=\tilde g^{\mu\nu}\partial_{\mu}\phi\,\partial_{\nu}\phi$ and
\begin{equation}
K(\phi)=\frac{F+3F'^{2}}{F^{2}} .
\label{eq:K}
\end{equation}
The Gauss--Bonnet term is not conformally invariant: in the Einstein frame it yields
$f(\phi)\tilde{\GB}$ together with additional derivative couplings of the scalar to the
curvature, whose explicit form we shall not need. The kinetic function~\eqref{eq:K} is positive
whenever $F>0$, so the scalar is not a ghost. For the family~\eqref{eq:F-family},
\begin{equation}
K_{X}=K_{\mathrm{Q}}-\gamma\,(1+12\beta)\,\beta^{2}\phi^{4}+O(\phi^{6}),\qquad
K_{\mathrm{Q}}=1+\frac{\beta}{4}(1+3\beta)\,\phi^{2}+\frac{\beta^{2}}{16}(1+6\beta)\,\phi^{4}
+O(\phi^{6}).
\label{eq:KX}
\end{equation}
This expansion shows that the completions are physically distinct theories rather than field
redefinitions of one another. The coefficient of $\tilde{\GB}$ in the Einstein frame is $f(\phi)$
for every completion, so a redefinition of the scalar that maps one completion onto another must
preserve $f$ and therefore reduce to $\phi\to\pm\phi$; by Eq.~\eqref{eq:KX}, the Einstein-frame
kinetic functions of two completions with different values of $\gamma$ then differ at fourth
order in $\phi$ unless $\beta=-1/12$.

The quantities that we compare between completions are frame independent. The horizon value
$\phi_{h}$ is a scalar. Since $F\to1$ at spatial infinity, where $\phi$ decays as $1/r$ (or
exponentially for $\mu>0$), the metrics $g_{\mu\nu}$ and $\tilde g_{\mu\nu}$ differ there only
at order $1/r^{2}$ and share the same ADM mass. The Hawking temperature is invariant under
conformal transformations that are regular and positive on the horizon~\cite{JacobsonKang1993},
which is the case here because $F(\phi_{h})>0$. Finally, the time coordinate of static observers
at infinity is common to both frames, and the map between the frames is invertible, so linear
perturbations and their frequencies correspond one to one; radial stability is therefore a
frame-independent property. Consequently, the domains of existence and of radial stability in
terms of $M/\sqrt{\alpha}$ at fixed couplings, which are the main objects of this work, are the
same in both frames.

In units $G=c=1$ the scalar field is dimensionless, the couplings have dimensions
$[\alpha]=L^{2}$, $[\mu]=L^{-1}$, and $[\lambda]=L^{-2}$, and $\beta$ and $\gamma$ are
dimensionless. The field equations are invariant under the rescaling $x^{\mu}\to s\,x^{\mu}$,
$\alpha\to s^{2}\alpha$, $\mu\to\mu/s$, $\lambda\to\lambda/s^{2}$, under which the action
scales as $S\to s^{2}S$ and masses and lengths scale as $M\to sM$. Physically distinct
configurations are therefore labeled by the dimensionless couplings $\beta, \gamma, \kappa_{\mu}\equiv\alpha\mu^{2}, \kappa_{\lambda}\equiv\alpha\lambda,$ and, along a branch at fixed couplings, by the normalized mass $\hM\equiv M/\sqrt{\alpha}$. When
we work in units of the horizon radius $r_{h}$, we also use $\zeta\equiv\alpha/r_{h}^{2}$.
Table~\ref{tab:conventions} relates these parameters to those of the works with which we compare
our results.

\begin{table}[t]
\caption{Relation between the parameters used here and those of other works. The dimensionless
couplings $\hat{\mu}\equiv\mu\sqrt{\eta}$ and $\hat{\lambda}\equiv\lambda\eta$ are those of
Ref.~\cite{Macedo2019}; $\lambda_{\mathrm{B}}$ and $\phi_{\mathrm{B}}$ denote the Gauss--Bonnet
coupling constant and the scalar field of Ref.~\cite{BlazquezSalcedo2018}.}
\label{tab:conventions}
\begin{ruledtabular}
\begin{tabular}{ll}
Reference & Relation to the present parameters\\
\hline
Refs.~\cite{Antoniou2021,Antoniou2022,Kleihaus2023} & identical $\alpha$ and $\beta$\\
Refs.~\cite{Silva2018,Macedo2019} & $\eta=4\alpha$,\quad $M/\sqrt{\eta}=\hM/2$,\quad
$\hat{\mu}=2\sqrt{\kappa_{\mu}}$,\quad $\hat{\lambda}=4\kappa_{\lambda}$\\
Ref.~\cite{BlazquezSalcedo2018} & $\lambda_{\mathrm{B}}^{2}=4\alpha$,\quad
$\phi_{\mathrm{B}}=\phi/2$,\quad $M/\lambda_{\mathrm{B}}=\hM/2$\\
\end{tabular}
\end{ruledtabular}
\end{table}

\section{Static scalarized black holes}
\label{sec:static}
\subsection{Reduced field equations}
\label{subsec:reduced}

We consider static, spherically symmetric configurations,
\begin{equation}
ds^{2}=-e^{2\delta(r)}N(r)\,dt^{2}+\frac{dr^{2}}{N(r)}+r^{2}\bigl(d\theta^{2}+\sin^{2}\theta\,d\varphi^{2}\bigr),
\qquad \phi=\phi(r),
\label{eq:ansatz}
\end{equation}
where $r$ is the areal radius. In this section a prime on $N$, $\delta$, or $\phi$ denotes
differentiation with respect to $r$, whereas $F'$, $f'$, and $V'$ continue to denote derivatives
with respect to $\phi$; radial derivatives of the composite functions $F(\phi(r))$ and
$f(\phi(r))$ are written as $dF/dr$ and $df/dr$. Substituting Eq.~\eqref{eq:ansatz} into the
action~\eqref{eq:action} and integrating over the angles gives
$S=\frac{1}{4}\int dt\,dr\,\mathcal{L}$ up to boundary terms, with
\begin{equation}
\mathcal{L}=e^{\delta}\Bigl\{2F\bigl[1-(rN)'\bigr]
+\bigl[r^{2}F'(\phi)-4(N-1)f'(\phi)\bigr]\phi'\,e^{-2\delta}\bigl(e^{2\delta}N\bigr)'
-\frac{1}{2}\,r^{2}N\phi'^{2}-r^{2}V\Bigr\}.
\label{eq:L1D}
\end{equation}
On the ansatz~\eqref{eq:ansatz} the curvature scalars are
\begin{equation}
\GB=\frac{4e^{-\delta}}{r^{2}}\Bigl[(N-1)\,e^{-\delta}\bigl(e^{2\delta}N\bigr)'\Bigr]',\qquad
R=\frac{2(1-N-2rN')}{r^{2}}-N''-3N'\delta'-2N\Bigl(\delta''+\delta'^{2}+\frac{2\delta'}{r}\Bigr).
\label{eq:GB-R-ansatz}
\end{equation}
The first expression displays the topological character of the Gauss--Bonnet invariant in four
dimensions: $\sqrt{-g}\,\GB$ is a total radial derivative. Consequently, the Gauss--Bonnet term
contributes to $\mathcal{L}$ only after an integration by parts, through the gradient
$df/dr=f'\phi'$, and with the weight $N-1$, which vanishes in flat space. The Ricci coupling, by
contrast, enters both through $F$ itself, which rescales the Einstein--Hilbert term, and through
its gradient $dF/dr=F'\phi'$.

Variation of $\mathcal{L}$ with respect to $N$ yields an equation that is algebraic in $\delta'$,
\begin{equation}
\delta'\Bigl[4rF+2r^{2}\frac{dF}{dr}-8(3N-1)\frac{df}{dr}\Bigr]
=r^{2}\phi'^{2}+2r^{2}\frac{d^{2}F}{dr^{2}}-8(N-1)\frac{d^{2}f}{dr^{2}} ,
\label{eq:delta-eq}
\end{equation}
variation with respect to $\delta$ yields an equation of first order in $N$,
\begin{align}
&4r^{2}N\frac{d^{2}F}{dr^{2}}+r^{2}N\phi'^{2}+2r^{2}V+2r^{2}N'\frac{dF}{dr}+4rFN'
+8rN\frac{dF}{dr}+4F(N-1)\nonumber\\
&\qquad-16N(N-1)\frac{d^{2}f}{dr^{2}}-8(3N-1)N'\frac{df}{dr}=0,
\label{eq:N-eq}
\end{align}
and variation with respect to $\phi$ reproduces the scalar equation $E_{\phi}=0$ of
Eq.~\eqref{eq:Ephi}, with $\Box\phi=e^{-\delta}r^{-2}\bigl(e^{\delta}r^{2}N\phi'\bigr)'$ and with
$R$ and $\GB$ given by Eq.~\eqref{eq:GB-R-ansatz}. We have verified that Eq.~\eqref{eq:L1D}
differs from the four-dimensional Lagrangian density evaluated on the ansatz by a total radial
derivative, and that its Euler--Lagrange equations coincide, up to nonvanishing factors, with
the components of Eqs.~\eqref{eq:Emn} and~\eqref{eq:Ephi}. For the minimally coupled case,
$F=1$ and $f=0$, Eq.~\eqref{eq:delta-eq} reduces to the familiar relation
$\delta'=r\phi'^{2}/4$.

\subsection{Boundary conditions and physical quantities}
\label{subsec:bc}

A nonextremal horizon at $r=r_{h}$ is a simple zero of $N$, and all fields are analytic there:
\begin{gather}
N=b_{1}(r-r_{h})+b_{2}(r-r_{h})^{2}+\dots,\qquad
\delta=\delta_{h}+\delta_{1}(r-r_{h})+\dots,\nonumber\\
\phi=\phi_{h}+\phi_{1}(r-r_{h})+\phi_{2}(r-r_{h})^{2}+\dots,
\label{eq:frobenius}
\end{gather}
with $b_{1}>0$. At spatial infinity the metric is asymptotically flat and the scalar field
vanishes. For a massless scalar,
\begin{equation}
N=1-\frac{2M}{r}+O(r^{-2}),\qquad \delta=O(r^{-2}),\qquad \phi=\frac{Q}{r}+O(r^{-2}),
\label{eq:asymptotics}
\end{equation}
which defines the ADM mass $M$ and the scalar charge $Q$; for $\mu>0$ the scalar decays as
$\phi\propto e^{-\mu r}r^{-1-\mu M}$. We fix the normalization of the time coordinate by
$\delta(\infty)=0$. The Hawking temperature and the Wald
entropy~\cite{Wald1993,IyerWald1994} are then
\begin{equation}
T=\frac{e^{\delta_{h}}\,b_{1}}{4\pi},\qquad S=\pi r_{h}^{2}F(\phi_{h})+4\pi f(\phi_{h}).
\label{eq:T-S}
\end{equation}
The Ricci coupling rescales the area term of the entropy, whereas the Gauss--Bonnet coupling adds a
term proportional to the Euler characteristic of the horizon cross section. At fixed couplings
our numerical solutions satisfy the first law $dM=T\,dS$ to a relative accuracy of $10^{-7}$ or
better, which provides a global check of the integration (Appendix~\ref{app:numerics}).

\subsection{Near-horizon regularity condition}
\label{subsec:regularity}

Inserting the expansion~\eqref{eq:frobenius} into the field equations, the lowest order fixes
\begin{equation}
b_{1}=\frac{2F_{h}-V_{h}r_{h}^{2}}{2F_{h}r_{h}+\mathcal{P}_{h}\phi_{1}},\qquad
\mathcal{P}_{h}\equiv r_{h}^{2}F'_{h}+4f'_{h},
\label{eq:b1}
\end{equation}
where $F_{h}\equiv F(\phi_{h})$, $F'_{h}\equiv F'(\phi_{h})$, and similarly for $f$ and $V$. At
the next order, the equations for $(\delta_{1},b_{2},\phi_{2})$ are linear with an identically
singular coefficient matrix. Their solvability condition is a quadratic equation for the horizon
gradient $\phi_{1}=\phi'(r_{h})$,
\begin{equation}
\mathcal{P}_{h}\Pi_{h}\,\phi_{1}^{2}+2r_{h}\mathcal{B}_{h}\,\phi_{1}-4F_{h}\mathcal{C}_{h}=0,
\label{eq:quadratic}
\end{equation}
with coefficients
\begin{align}
\Pi_{h}={}&2F_{h}\mathcal{Q}_{h}-r_{h}^{4}V_{h}\bigl(F_{h}+F_{h}'^{2}\bigr)+16V_{h}f_{h}'^{2}
-F_{h}r_{h}^{2}V'_{h}\mathcal{P}_{h},\nonumber\\
\mathcal{B}_{h}={}&2F_{h}^{2}\bigl(\mathcal{Q}_{h}-r_{h}^{2}V'_{h}\mathcal{P}_{h}\bigr)
+F_{h}V_{h}\bigl[\mathcal{P}_{h}\bigl(r_{h}^{2}F'_{h}+12f'_{h}\bigr)-F_{h}r_{h}^{4}\bigr]
-2r_{h}^{2}f'_{h}V_{h}^{2}\mathcal{P}_{h},\nonumber\\
\mathcal{C}_{h}={}&F_{h}^{2}r_{h}^{4}V'_{h}-12F_{h}^{2}f'_{h}-2F_{h}F'_{h}V_{h}r_{h}^{4}
+4F_{h}V_{h}f'_{h}r_{h}^{2}-r_{h}^{4}V_{h}^{2}f'_{h},
\label{eq:quad-coeffs}
\end{align}
where
\begin{equation}
\mathcal{Q}_{h}=r_{h}^{2}\bigl(F_{h}+3F_{h}'^{2}\bigr)+12F'_{h}f'_{h},\qquad
\mathcal{S}_{h}=r_{h}^{4}\bigl(F_{h}+3F_{h}'^{2}\bigr)-12r_{h}^{2}F'_{h}f'_{h}-96f_{h}'^{2},
\label{eq:QS}
\end{equation}
which satisfy 
\begin{equation}
  \mathcal{S}_{h}=r_{h}^{2}\mathcal{Q}_{h}-24\mathcal{P}_{h}f'_{h}.
\end{equation} 
The condition
involves only the values of $F$, $F'$, $f'$, $V$, and $V'$ on the horizon; the second
derivatives of the coupling functions drop out. The discriminant of Eq.~\eqref{eq:quadratic}
equals $4(2F_{h}-V_{h}r_{h}^{2})^{2}\Delta_{h}$, with
\begin{equation}
\Delta_{h}=F_{h}^{2}\mathcal{Q}_{h}\mathcal{S}_{h}
+8F_{h}f'_{h}\mathcal{P}_{h}\bigl(F_{h}r_{h}^{4}-6f'_{h}\mathcal{P}_{h}\bigr)V_{h}
+4r_{h}^{2}f_{h}'^{2}\mathcal{P}_{h}^{2}V_{h}^{2}+12F_{h}^{2}r_{h}^{2}f'_{h}\mathcal{P}_{h}^{2}V'_{h}.
\label{eq:Delta}
\end{equation}
A regular horizon therefore exists only if $\Delta_{h}\geq0$. The physical root, which vanishes
with $\phi_{h}$ and is continuously connected to the Schwarzschild solution, can be written as
\begin{equation}
\phi_{1}=\frac{4F_{h}\,\mathcal{C}_{h}}{r_{h}\mathcal{B}_{h}+\bigl(2F_{h}-V_{h}r_{h}^{2}\bigr)\sqrt{\Delta_{h}}} ,
\label{eq:root}
\end{equation}
a form that remains finite at $\mathcal{P}_{h}=0$, where the leading coefficient of
Eq.~\eqref{eq:quadratic} vanishes and the textbook expression for the root becomes
indeterminate. For a massless scalar these expressions simplify to
\begin{equation}
\Delta_{h}\big|_{V=0}=F_{h}^{2}\mathcal{Q}_{h}\mathcal{S}_{h},\qquad
\phi_{1}\big|_{V=0}=-\frac{24F_{h}f'_{h}}{r_{h}\mathcal{Q}_{h}+\sqrt{\mathcal{Q}_{h}\mathcal{S}_{h}}} .
\label{eq:root-massless}
\end{equation}

Equations~\eqref{eq:quadratic}--\eqref{eq:root-massless} contain the known special cases. For
$F=1$ and $V=0$ one has $\mathcal{Q}_{h}=r_{h}^{2}$ and $\mathcal{S}_{h}=r_{h}^{4}-96f_{h}'^{2}$,
and the condition becomes $r_{h}^{4}\geq96f'(\phi_{h})^{2}$, the regularity condition for a
general Gauss--Bonnet coupling~\cite{DonevaYazadjiev2018,AntoniouBakopoulosKanti2018}; for the
quadratic coupling it reads $r_{h}^{4}\geq96\alpha^{2}\phi_{h}^{2}$, in agreement with
Ref.~\cite{Silva2018}. Our condition also reproduces the result of Ref.~\cite{Antoniou2021} for
quadratic Ricci and Gauss--Bonnet couplings without a potential, and that of
Ref.~\cite{Macedo2019} for $F=1$ with a potential; we have verified both reductions
symbolically.

The massless form admits a transparent interpretation. By Eq.~\eqref{eq:K},
$r_{h}^{4}(F_{h}+3F_{h}'^{2})=r_{h}^{4}F_{h}^{2}K(\phi_{h})$, so that
\begin{equation}
\mathcal{S}_{h}=r_{h}^{4}F_{h}^{2}K(\phi_{h})-12r_{h}^{2}F'_{h}f'_{h}-96f_{h}'^{2}.
\label{eq:S-einstein}
\end{equation}
The first term is set by the Einstein-frame kinetic function of the scalar on the horizon and is
positive. The last term originates from the Gauss--Bonnet coupling and is negative: a regular
horizon requires the kinetic term, weighted by $r_{h}^{4}$, to dominate the square of the
horizon gradient $f'(\phi_{h})$ of the Gauss--Bonnet coupling. For $\alpha,\beta>0$ the mixed
term, $-12r_{h}^{2}F'_{h}f'_{h}=6\alpha\beta r_{h}^{2}\phi_{h}^{2}+O(\phi_{h}^{4})$, is positive at
leading order, so that the Ricci coupling relaxes the regularity bound. When $\Delta_{h}$
decreases to zero along a branch, the two roots of Eq.~\eqref{eq:quadratic} merge; beyond this
point no real horizon gradient exists, and the family of regular black holes terminates. Along all
massless branches constructed in this work $\mathcal{Q}_{h}>0$, so that this discriminant closure
occurs at $\mathcal{S}_{h}=0$. In Sec.~\ref{sec:existence} we show that discriminant closure
terminates every branch that we have computed.

\subsection{Horizon curvature}
\label{subsec:curvature}

In the static orthonormal frame, the nonvanishing components of the Riemann tensor on a regular
horizon are
\begin{equation}
R^{\hat t\hat r}{}_{\hat t\hat r}=-\frac{X}{2},\qquad
R^{\hat t\hat\theta}{}_{\hat t\hat\theta}=R^{\hat r\hat\theta}{}_{\hat r\hat\theta}=-\frac{b_{1}}{2r_{h}},\qquad
R^{\hat\theta\hat\varphi}{}_{\hat\theta\hat\varphi}=\frac{1}{r_{h}^{2}},\qquad
X\equiv2b_{2}+3b_{1}\delta_{1},
\label{eq:riemann-h}
\end{equation}
together with those related to them by spherical symmetry, so that
\begin{equation}
R_{h}=-X-\frac{4b_{1}}{r_{h}}+\frac{2}{r_{h}^{2}},\qquad
\GB_{h}=\frac{4}{r_{h}^{2}}\bigl(b_{1}^{2}-X\bigr),\qquad
K_{h}\equiv R_{\mu\nu\rho\sigma}R^{\mu\nu\rho\sigma}\big|_{h}=X^{2}+\frac{4b_{1}^{2}}{r_{h}^{2}}+\frac{4}{r_{h}^{4}} .
\label{eq:invariants-h}
\end{equation}
On the horizon the scalar equation reduces to $b_{1}\phi_{1}+F'_{h}R_{h}-V'_{h}+f'_{h}\GB_{h}=0$,
which is linear in $X$ and gives
\begin{equation}
X=\frac{r_{h}^{2}}{\mathcal{P}_{h}}\Bigl[b_{1}\phi_{1}-V'_{h}
+F'_{h}\Bigl(\frac{2}{r_{h}^{2}}-\frac{4b_{1}}{r_{h}}\Bigr)+\frac{4f'_{h}b_{1}^{2}}{r_{h}^{2}}\Bigr].
\label{eq:X}
\end{equation}
Together with Eqs.~\eqref{eq:b1} and~\eqref{eq:root}, this expresses every curvature invariant on
the horizon in closed form in terms of $r_{h}$, $\phi_{h}$, and the couplings. The apparent
singularity at $\mathcal{P}_{h}=0$ is removable: on the physical root the bracket in
Eq.~\eqref{eq:X} vanishes there as well, and $X$ is finite and continuous. For the Schwarzschild
solution, $b_{1}=1/r_{h}$ and $X=-2/r_{h}^{2}$, and Eq.~\eqref{eq:invariants-h} gives $R_{h}=0$
and $K_{h}=\GB_{h}=12/r_{h}^{4}$.

These closed forms have a direct geometric consequence. At the discriminant closure,
$\phi_{1}$ and $b_{1}$ remain finite, and so do $X$ and all curvature invariants whenever
$\mathcal{P}_{h}\neq0$: the branches of scalarized black holes end at finite horizon curvature,
not at a curvature singularity. The invariants can nevertheless become large. Along the branch of
the quadratic truncation with $\beta=1.2$, for example, the Kretschmann scalar grows from its
Schwarzschild value $12/r_{h}^{4}$ at the threshold to about $436/r_{h}^{4}$ at
$\phi_{h}=0.85$, close to the end of the branch, where $R_{h}\simeq16/r_{h}^{2}$. The
effective-field-theory interpretation of the action becomes questionable in this regime, a point
to which we return in Sec.~\ref{sec:discussion}.

\section{Weakly nonlinear analysis: exact bifurcation laws}
\label{sec:wnl}
Near the threshold~\eqref{eq:threshold} the scalarized branch can be constructed as a series in
the horizon value of the scalar. The coefficient of the leading correction to the mass decides
whether the branch leaves the Schwarzschild family toward larger or smaller values of
$M/\sqrt{\alpha}$ and, as we show in Sec.~\ref{sec:stability}, whether the black holes near the
threshold are radially stable. In this section we compute this coefficient exactly for a
massless scalar, determine its dependence on a quartic self-interaction, and extend the analysis
to a massive scalar.

\subsection{Amplitude expansion}
\label{subsec:expansion}

We work in units $r_{h}=1$, in which the Gauss--Bonnet coupling is measured by
$\zeta=\alpha/r_{h}^{2}$, and expand at fixed dimensionless couplings in the amplitude
$\varepsilon\equiv\phi_{h}$. Since the field equations are invariant under $\phi\to-\phi$, the
expansion takes the form
\begin{gather}
\phi=\varepsilon\phi_{1}+\varepsilon^{3}\phi_{3}+O(\varepsilon^{5}),\qquad
N=N_{0}+\varepsilon^{2}N_{2}+O(\varepsilon^{4}),\nonumber\\
\delta=\varepsilon^{2}\delta_{2}+O(\varepsilon^{4}),\qquad
\zeta=\zeta_{0}+\varepsilon^{2}\zeta_{2}+O(\varepsilon^{4}),
\label{eq:expansion}
\end{gather}
with $N_{0}=1-1/r$, $\phi_{1}(1)=1$, $\phi_{3}(1)=0$, and $N_{2}(1)=0$, so that the horizon stays
at $r=1$ and $\phi_{h}=\varepsilon$ exactly. Writing $N_{2}\to-2M_{2}/r$ at infinity, the mass is
$M=\frac{1}{2}+\varepsilon^{2}M_{2}+O(\varepsilon^{4})$, and the normalized mass along the branch
becomes
\begin{equation}
\hM=\hM_{c}+k_{2}\,\phi_{h}^{2}+O(\phi_{h}^{4}),\qquad
k_{2}=\frac{1}{2\sqrt{\zeta_{0}}}\Bigl(2M_{2}-\frac{\zeta_{2}}{2\zeta_{0}}\Bigr),
\label{eq:k2-def}
\end{equation}
with $\hM_{c}=1/(2\sqrt{\zeta_{0}})$. The coefficient $k_{2}$ fixes the direction of the
bifurcation. Since the Schwarzschild solution is linearly unstable for $\hM<\hM_{c}$, a branch
with $k_{2}>0$ initially coexists with stable Schwarzschild black holes (a subcritical
bifurcation), whereas a branch with $k_{2}<0$ extends into the region in which the Schwarzschild
solution is unstable (a supercritical bifurcation). Because $\gamma$ enters the field equations
only at fourth order in the amplitude (Sec.~\ref{subsec:completions}), the results of this
section apply to all completions alike.

\subsection{First order: the zero mode and the threshold}
\label{subsec:zero-mode}

At first order, the scalar equation reduces to the static zero-mode equation on the
Schwarzschild background,
\begin{equation}
\mathsf{L}[\phi_{1}]\equiv\bigl(r^{2}N_{0}\phi_{1}'\bigr)'+\frac{12\zeta_{0}}{r^{4}}\,\phi_{1}=0 .
\label{eq:zero-mode}
\end{equation}
In terms of $v=1-1/r\in[0,1)$ it becomes
\begin{equation}
  v\,\partial_{v}^{2}\phi_{1}+\partial_{v}\phi_{1}+3A(1-v)^{2}\phi_{1}=0, \qquad \text{with} \qquad A=4\zeta_{0}=\alpha/M^{2}.
\end{equation} 
Its only singular points are $v=0$ and $v=\infty$, so the solution
that is regular on the horizon is an entire function of $v$,
\begin{equation}
\phi_{1}=\sum_{n\geq0}c_{n}v^{n},\qquad c_{0}=1,\qquad
(n+1)^{2}c_{n+1}=-3A\,(c_{n}-2c_{n-1}+c_{n-2}),
\label{eq:series}
\end{equation}
with $c_{-1}=c_{-2}=0$; the series converges on the whole exterior, including spatial infinity
($v=1$). A static zero mode that decays at infinity therefore exists if and only if
$\phi_{1}(v=1)=\sum_{n}c_{n}(A)=0$. The smallest positive root of this condition reproduces the
threshold $A_{00}=0.7256128492627$ of Eq.~\eqref{eq:threshold} and provides a third, independent
determination of it. At large $r$, $\phi_{1}\simeq Q_{1}/r$. The operator $\mathsf{L}$ is
self-adjoint on $[1,\infty)$ for functions that are regular on the horizon and decay at infinity,
since the boundary terms $\bigl[r^{2}N_{0}(u\,w'-w\,u')\bigr]_{1}^{\infty}$ vanish for such
functions.

\subsection{Second order: back-reaction and an exact Ricci term}
\label{subsec:second-order}

At second order the scalar field back-reacts on the geometry through $f=\varepsilon^{2}f_{2}$ and
$F=1+\varepsilon^{2}F_{2}+O(\varepsilon^{4})$, where $f_{2}=\zeta_{0}\phi_{1}^{2}/2$ and
$F_{2}=-\beta\phi_{1}^{2}/4$ are regarded as functions of $r$. Equation~\eqref{eq:delta-eq}
determines $\delta_{2}'$ algebraically, and Eq.~\eqref{eq:N-eq} becomes
\begin{equation}
4\bigl(rN_{2}\bigr)'+r^{2}N_{0}\phi_{1}'^{2}-16N_{0}(N_{0}-1)f_{2}''-8(3N_{0}-1)N_{0}'f_{2}'
+\bigl(4r^{2}N_{0}F_{2}'-2F_{2}\bigr)'=0 .
\label{eq:N2-eq}
\end{equation}
The entire dependence on the Ricci coupling is contained in the last term, which is a total
derivative. Integrating from the horizon, where $rN_{2}=N_{0}=0$, to infinity, where
$rN_{2}\to-2M_{2}$ and $\phi_{1}\to0$, gives
\begin{equation}
M_{2}=\frac{1}{8}\int_{1}^{\infty}\Bigl[r^{2}N_{0}\phi_{1}'^{2}-16N_{0}(N_{0}-1)f_{2}''
-8(3N_{0}-1)N_{0}'f_{2}'\Bigr]dr-\frac{\beta}{16},
\label{eq:M2}
\end{equation}
where the integral evaluates to $0.0907016061$. The Ricci coupling thus contributes to the
second-order mass only through the value $F_{2}(r_{h})=-\beta\phi_{h}^{2}/4$ of the coupling on the
horizon. Restoring units, it lowers the mass of a black hole with given $r_{h}$ and $\phi_{h}$ by
exactly $\beta\phi_{h}^{2}r_{h}/16$.

\subsection{Third order: solvability and the tilt law}
\label{subsec:tilt}

At third order the scalar equation takes the form $\mathsf{L}[\phi_{3}]+S_{3}=0$, where the
source $S_{3}$ is built from $\phi_{1}$, $N_{2}$, $\delta_{2}$, and $\zeta_{2}$ and depends on
$\zeta_{2}$ only through the term $12\zeta_{2}\phi_{1}/r^{4}$. Multiplying by $\phi_{1}$,
integrating over the exterior, and using the self-adjointness of $\mathsf{L}$ together with
$\mathsf{L}[\phi_{1}]=0$ yields the solvability condition
\begin{equation}
\int_{1}^{\infty}\phi_{1}S_{3}\,dr=0 ,
\label{eq:fredholm}
\end{equation}
which is linear in $\zeta_{2}$, with coefficient
$I_{z}\equiv\int_{1}^{\infty}12\phi_{1}^{2}r^{-4}\,dr=1.7441821390$, and therefore determines it
uniquely.

The dependence of $k_{2}$ on the Ricci coupling follows from power counting. In the scalar
equation~\eqref{eq:Ephi}, $\beta$ enters only through the term $F'(\phi)R$, and $R$ vanishes on
the Schwarzschild background. At third order this term contributes $-\frac{\beta}{2}\phi_{1}R^{(2)}$,
where the second-order Ricci scalar $R^{(2)}$ is at most linear in $\beta$, as are $N_{2}$ and
$\delta_{2}$. Hence $\zeta_{2}$ is exactly a quadratic polynomial in $\beta$, and so, by
Eqs.~\eqref{eq:k2-def} and~\eqref{eq:M2}, is $k_{2}$. Evaluating the integrals with the zero
mode~\eqref{eq:series}, we find
\begin{equation}
k_{2}(\beta)=0.2949551634-0.0608213273\,\beta-0.1824639822\,\beta^{2}.
\label{eq:tilt-law}
\end{equation}
Table~\ref{tab:tilt} lists the underlying values of $\zeta_{2}$ and $M_{2}$. The linear
coefficient contains the exact contribution $-\hM_{c}/8=-0.1467431$ of the horizon term in
Eq.~\eqref{eq:M2}. The negative quadratic coefficient originates entirely from the term
$F'R^{(2)}$, that is, from the action of the Ricci coupling on the part of the second-order Ricci
curvature that the Ricci coupling itself induces. The tilt changes sign at
\begin{equation}
\btilt=1.1156319 ,
\label{eq:beta-tilt}
\end{equation}
so that the bifurcation is subcritical for $0\leq\beta<\btilt$ and supercritical for
$\beta>\btilt$. The second root, $\beta=-1.4490$, marks the corresponding reversal for a negative
Ricci coupling, consistent with the observation of Ref.~\cite{Antoniou2021} that a critical value
exists also for $\beta<0$.

\begin{table}[t]
\caption{Second-order coefficients of the amplitude expansion for a massless scalar, in units
$r_{h}=1$: the coupling correction $\zeta_{2}$, the mass correction $M_{2}$, and the tilt
$k_{2}=d\hM/d\phi_{h}^{2}$ at the threshold. $M_{2}$ is linear in $\beta$ with slope $-1/16$,
while $\zeta_{2}$ and $k_{2}$ are quadratic; the entries at $\beta=1.1$, $1.2$, and $3$ are
reproduced by Eq.~\eqref{eq:tilt-law} to better than $10^{-11}$.}
\label{tab:tilt}
\begin{ruledtabular}
\begin{tabular}{crrr}
$\beta$ & $\zeta_{2}$ & $M_{2}$ & $k_{2}$\\
\hline
$0$ & $-0.0253413702$ & $0.0907016061$ & $0.2949551634$\\
$1.0$ & $0.0044949271$ & $0.0282016062$ & $0.0516698540$\\
$1.1$ & $0.0136814926$ & $0.0219516062$ & $0.0072702850$\\
$1.2$ & $0.0239958647$ & $0.0157016062$ & $-0.0407785637$\\
$2.0$ & $0.1471118752$ & $-0.0342983938$ & $-0.5565434198$\\
$3.0$ & $0.4025094739$ & $-0.0967983937$ & $-1.5296846579$\\
\end{tabular}
\end{ruledtabular}
\end{table}

The tilt law agrees with our nonlinear solutions. Direct shooting for the sign change of the tilt
gives $\btilt=1.11562$, and fits to near-threshold solutions give $\zeta_{2}=-0.025345$ and
$0.004493$ at $\beta=0$ and $1$, compared with $-0.0253414$ and $0.0044949$ from the expansion.
Earlier estimates, $\beta\approx1.22$ from the tilt of the charge--mass
curves~\cite{Antoniou2021} and $\beta\approx1.15$--$1.2$ for the appearance of radially stable
solutions~\cite{Antoniou2022}, were obtained at finite amplitude. For $\beta$ slightly above
$\btilt$ the branch leaves the threshold in the supercritical direction but turns back at a
nearby turning point (Sec.~\ref{sec:stability}), so that its initial direction is difficult to
resolve numerically.

\subsection{Quartic self-interaction}
\label{subsec:quartic}

For $V=\lambda\phi^{4}/2$ the potential is of fourth order in the amplitude and does not affect the
geometry at second order. At third order it adds $-2\lambda r^{2}\phi_{1}^{3}$ to $S_{3}$, with
$\lambda$ in units of $r_{h}^{-2}$. The solvability condition~\eqref{eq:fredholm} then shifts
$\zeta_{2}$ by $2\lambda J_{4}/I_{z}$, where $J_{4}\equiv\int_{1}^{\infty}r^{2}\phi_{1}^{4}\,dr
=0.2445619248$. Since $\kappa_{\lambda}=\alpha\lambda=\zeta_{0}\lambda$ at this order, the tilt law
generalizes to
\begin{equation}
k_{2}(\beta,\kappa_{\lambda})=k_{2}(\beta)-c_{\lambda}\,\kappa_{\lambda},\qquad
c_{\lambda}=\frac{J_{4}}{2\zeta_{0}^{5/2}I_{z}}=5.0021254566,
\label{eq:quartic-law}
\end{equation}
exactly in $\beta$ and $\kappa_{\lambda}$. The effect has a transparent physical origin: a
repulsive self-interaction saturates the growth of the tachyonic instability, so that a larger
coupling $\alpha/r_{h}^{2}$ is required to support a scalar of given amplitude, and the branch is
pushed toward smaller $M/\sqrt{\alpha}$, into the supercritical direction. The bifurcation
becomes supercritical for
\begin{equation}
\kappa_{\lambda}>\kappa_{\lambda}^{*}(\beta)=\frac{k_{2}(\beta)}{c_{\lambda}},
\label{eq:kappa-star}
\end{equation}
with $\kappa_{\lambda}^{*}=0.05896597$, $0.04376710$, and $0.01032958$ for $\beta=0$, $0.5$, and
$1$, and $\kappa_{\lambda}^{*}\to0$ as $\beta\to\btilt$. The Ricci coupling and the
self-interaction therefore act cooperatively, and a modest self-interaction suffices even in the
absence of a Ricci coupling. Nonlinear solutions confirm Eq.~\eqref{eq:quartic-law}: for
$\beta=0$ they give $k_{2}=+0.0448527$ at $\kappa_{\lambda}=0.05$ and $k_{2}=-0.2052306$ at
$\kappa_{\lambda}=0.1$, compared with $+0.0448489$ and $-0.2052574$ from the law.

\subsection{Massive scalar}
\label{subsec:massive-tilt}

A mass term modifies the linear problem: the threshold moves, the zero mode acquires a Yukawa
tail, and the integrals above must be evaluated with the massive zero mode. The structure of the
argument is unchanged, however. The quartic term still enters only through the third-order
source, so that $k_{2}$ remains exactly linear in $\kappa_{\lambda}$ at fixed $\kappa_{\mu}$. We
have determined the coefficients from nonlinear solutions near the threshold. For $\beta=0$ and
$\kappa_{\mu}=6.25\times10^{-4}$, for which $\hM_{c}=1.149333$,
\begin{equation}
k_{2}=0.30598936-3.88610270\,\kappa_{\lambda},
\label{eq:massive-law}
\end{equation}
and the bifurcation becomes supercritical for $\kappa_{\lambda}>\kappa_{\lambda}^{*}=0.078739$.
The mass makes the bifurcation more strongly subcritical and weakens the effect of the
self-interaction, raising the critical self-interaction by about one third relative to the
massless value. In the notation of Ref.~\cite{Macedo2019}, these couplings correspond to
$\hat{\mu}=0.05$ and $\hat{\lambda}^{*}=0.315$. In Sec.~\ref{sec:stability} we confirm this
threshold independently by counting unstable radial modes, and in Sec.~\ref{sec:discussion} we
compare it with the results of Ref.~\cite{Macedo2019}.

Equations~\eqref{eq:tilt-law}, \eqref{eq:quartic-law}, and~\eqref{eq:massive-law} thus provide
explicit criteria for the direction of the bifurcation. As we show in Sec.~\ref{sec:stability},
this direction determines the radial stability of the scalarized black holes near the threshold,
in accordance with the exchange of stability familiar from bifurcation theory: supercritical
branches are radially stable near the threshold, and subcritical branches are unstable.

\section{Domains of existence}
\label{sec:existence}
For fixed couplings $(\beta,\gamma,\kappa_{\mu},\kappa_{\lambda})$ the fundamental branch of
scalarized black holes, that is, the nodeless branch that bifurcates from the Schwarzschild family
at the threshold of Sec.~\ref{subsec:threshold}, is a one-parameter family, which we label by the
horizon value $\phi_{h}$ of the scalar. We denote by $\mathcal{E}_{X}$ the domain of existence of
the completion $X$, that is, the set of values of the normalized mass $\hM=M/\sqrt{\alpha}$ attained
along this branch. In this section we determine how the branches terminate and compare the domains
of the quadratic truncation and of the exponential completion; the completions R and M are
discussed in Sec.~\ref{sec:mechanism}.

\subsection{Construction and termination of the branches}
\label{subsec:termination}

We construct the solutions by integrating the field equations outward from $r=r_{h}+\epsilon$ with
the horizon data of Sec.~\ref{subsec:regularity}, determining $\zeta=\alpha/r_{h}^{2}$ by the
requirement that the scalar vanish at infinity (or decay exponentially for $\mu>0$), and
extrapolating in the offset $\epsilon$. Near the ends of the branches we use pseudo-arclength
continuation. The numerical methods and the error control are described in
Appendix~\ref{app:numerics}; the endpoints of the domains are accurate to about $10^{-5}$ in $\hM$.

Every branch of the completions Q and E that we have followed to its end terminates by
discriminant closure (Sec.~\ref{subsec:regularity}). This comprises, for each completion, the
massless branches at 17 values of $\beta$ between $0.5$ and $5$, the common branch at $\beta=0$,
and all 24 branches of the massive and self-interacting sector discussed in
Sec.~\ref{subsec:massive-sector}. In all cases the branch is a single-valued function of $\phi_{h}$
up to its end, and the quantity $2F_{h}-V_{h}r_{h}^{2}$, whose vanishing would signal an extremal
horizon [cf.\ Eq.~\eqref{eq:b1}], remains above $0.44$. For a massless scalar, $\mathcal{Q}_{h}$
remains positive and the ratio $\mathcal{S}_{h}/\mathcal{Q}_{h}$ decreases to zero at the end, with
$\mathcal{S}_{h}/\mathcal{Q}_{h}\leq1.6\times10^{-4}$ on the last computed solutions; in the massive
sector $\Delta_{h}$ vanishes at the end in the same way. The branches thus end at a regular horizon
with finite curvature (Sec.~\ref{subsec:curvature}), beyond which no regular black hole with the
required asymptotics exists. The curvature at the end can nevertheless be large, particularly for
large $\beta$: for the quadratic truncation at $\beta=5$, where $\mathcal{P}_{h}$ becomes small
close to the end, the Kretschmann scalar on the horizon of the last computed solution reaches
$1.5\times10^{6}/r_{h}^{4}$.

For $\beta=0$ all completions reduce to $F=1$, and for $\mu=\lambda=0$ the theory reduces to
scalar--Gauss--Bonnet gravity with a quadratic coupling. The fundamental branch then spans
$\hM\in[1.173944,\,1.273626]$, that is, $\eta/M^{2}\in[2.4659,\,2.9025]$ in the normalization of
Ref.~\cite{Silva2018}, and it ends exactly where the bound $r_{h}^{4}\geq96\alpha^{2}\phi_{h}^{2}$
becomes saturated (to $4\times10^{-5}$ on the last computed solution). The endpoint corresponds to
$M/\lambda_{\mathrm{B}}=0.6368$ in the normalization of Ref.~\cite{BlazquezSalcedo2018}, in
agreement to $0.13\%$ with the edge $M/\lambda_{\mathrm{B}}\approx0.636$ of the fundamental branch
reported there, as quoted in Ref.~\cite{MyungZou2019}. Reference~\cite{Silva2018} quotes
$\eta/M^{2}\approx2.53$ for the lower edge of the branch. That point lies at about $91\%$ of the
final value of $\phi_{h}$, where the regularity bound is not yet saturated
($96\zeta^{2}\phi_{h}^{2}\approx0.84$), and the solutions with $2.466<\eta/M^{2}<2.53$ satisfy the
regularity condition, the field equations, and the first law, the latter to $2\times10^{-7}$.

\subsection{Massless completions: transition structure}
\label{subsec:massless-domains}

Table~\ref{tab:domains} lists the domains of existence of the completions Q and E for a massless
scalar at 18 values of $\beta$. On subcritical branches ($\beta<\btilt$) the scalarized black
holes are heavier than the threshold and the domains have the form $[\hM_{c},\hM_{\max}]$. On
supercritical branches ($\beta>\btilt$) the branch first descends below the threshold to a minimum
of $\hM$ and then rises again; the lower edge of the domain is this minimum, and the upper edge is
$\hM_{c}$ or the endpoint of the branch, whichever is larger. Comparing the two completions reveals
a sequence of four regimes separated by three transitions.

\begin{table}[t]
\caption{Domains of existence of the fundamental branch for a massless scalar: lower and upper
edges of $\mathcal{E}_{X}$ in $\hM=M/\sqrt{\alpha}$ and the horizon value of the scalar at the end
of the branch, for the quadratic truncation (Q) and the exponential completion (E). The last
column states which domain contains the other. The edges are accurate to about $10^{-5}$; the
threshold is $\hM_{c}=1.173944$.}
\label{tab:domains}
\begin{ruledtabular}
\begin{tabular}{cccccccc}
& \multicolumn{3}{c}{Quadratic truncation (Q)} & \multicolumn{3}{c}{Exponential completion (E)} &\\
$\beta$ & $\hM_{\min}$ & $\hM_{\max}$ & $\phi_{h}^{\mathrm{end}}$ & $\hM_{\min}$ & $\hM_{\max}$ &
$\phi_{h}^{\mathrm{end}}$ & Relation\\
\hline
$0$ & $1.173944$ & $1.273626$ & $0.5890$ & $1.173944$ & $1.273626$ & $0.5890$ & $\mathrm{Q}=\mathrm{E}$\\
$0.5$ & $1.173944$ & $1.268850$ & $0.6687$ & $1.173944$ & $1.268563$ & $0.6623$ & $\mathrm{Q}\supsetneq\mathrm{E}$\\
$0.6$ & $1.173944$ & $1.263832$ & $0.6910$ & $1.173944$ & $1.263718$ & $0.6799$ & $\mathrm{Q}\supsetneq\mathrm{E}$\\
$0.7$ & $1.173944$ & $1.256848$ & $0.7157$ & $1.173944$ & $1.257287$ & $0.6980$ & $\mathrm{E}\supsetneq\mathrm{Q}$\\
$0.8$ & $1.173944$ & $1.247539$ & $0.7430$ & $1.173944$ & $1.249191$ & $0.7160$ & $\mathrm{E}\supsetneq\mathrm{Q}$\\
$0.9$ & $1.173944$ & $1.235536$ & $0.7726$ & $1.173944$ & $1.239421$ & $0.7335$ & $\mathrm{E}\supsetneq\mathrm{Q}$\\
$1$ & $1.173944$ & $1.220523$ & $0.8038$ & $1.173944$ & $1.228057$ & $0.7498$ & $\mathrm{E}\supsetneq\mathrm{Q}$\\
$\btilt$ & $1.173944$ & $1.199226$ & $0.8406$ & $1.173944$ & $1.213151$ & $0.7667$ & $\mathrm{E}\supsetneq\mathrm{Q}$\\
$1.2$ & $1.168503$ & $1.181113$ & $0.8667$ & $1.171644$ & $1.201263$ & $0.7774$ & neither\\
$1.25$ & $1.161288$ & $1.173944$ & $0.8815$ & $1.168499$ & $1.193896$ & $0.7830$ & neither\\
$1.3$ & $1.152026$ & $1.173944$ & $0.8955$ & $1.164349$ & $1.186328$ & $0.7881$ & neither\\
$1.35$ & $1.141266$ & $1.173944$ & $0.9086$ & $1.159398$ & $1.178593$ & $0.7927$ & neither\\
$1.4$ & $1.129435$ & $1.173944$ & $0.9207$ & $1.153809$ & $1.173944$ & $0.7967$ & $\mathrm{Q}\supsetneq\mathrm{E}$\\
$1.45$ & $1.116859$ & $1.173944$ & $0.9316$ & $1.147721$ & $1.173944$ & $0.8003$ & $\mathrm{Q}\supsetneq\mathrm{E}$\\
$1.5$ & $1.103793$ & $1.173944$ & $0.9415$ & $1.141222$ & $1.173944$ & $0.8033$ & $\mathrm{Q}\supsetneq\mathrm{E}$\\
$2$ & $0.973307$ & $1.173944$ & $0.9838$ & $1.067953$ & $1.173944$ & $0.8123$ & $\mathrm{Q}\supsetneq\mathrm{E}$\\
$3$ & $0.786334$ & $1.173944$ & $0.9313$ & $0.935550$ & $1.173944$ & $0.7750$ & $\mathrm{Q}\supsetneq\mathrm{E}$\\
$5$ & $0.594291$ & $1.173944$ & $0.7887$ & $0.763023$ & $1.173944$ & $0.6777$ & $\mathrm{Q}\supsetneq\mathrm{E}$\\
\end{tabular}
\end{ruledtabular}
\end{table}

(i) At $\beta=0$ the two completions coincide. (ii) For $0<\beta<\beta_{\times}$ both branches are
subcritical and the quadratic truncation extends marginally further, by $2.9\times10^{-4}$ in
$\hM$ at $\beta=0.5$ and $1.1\times10^{-4}$ at $\beta=0.6$; these margins exceed the uncertainty of
the edges by about an order of magnitude. (iii) The upper edges cross at
$\beta_{\times}\simeq0.63$ (quadratic interpolation of the margins, with an uncertainty of about
$0.01$), beyond which the exponential completion extends further, by up to $1.4\times10^{-2}$ at
$\btilt$. (iv) For $\btilt<\beta<\beta_{c2}$ both branches are supercritical, and neither domain
contains the other: the quadratic truncation reaches lower masses, whereas the branch of the
exponential completion ends above the threshold and therefore reaches higher masses. The endpoint
of the latter returns to the threshold at $\beta_{c2}=1.380\pm0.002$. (v) For $\beta>\beta_{c2}$
the domain of the quadratic truncation strictly contains that of the exponential completion, with
margins that grow with $\beta$: $0.024$ at $\beta=1.4$, $0.037$ at $1.5$, $0.095$ at $2$, $0.149$
at $3$, and $0.169$ at $5$. The transition at $\btilt$ is exact, whereas $\beta_{\times}$ and
$\beta_{c2}$ are obtained by interpolation.

Two further regularities are apparent. First, for every $\beta>0$ the branch of the exponential
completion terminates at a smaller horizon value of the scalar than that of the quadratic
truncation. Second, the differences between the two domains are small for small $\beta$ and grow
rapidly once the branches become supercritical. Both observations have a simple origin, which we
identify in Sec.~\ref{sec:mechanism}.

\subsection{Massive and self-interacting sector}
\label{subsec:massive-sector}

To test whether this structure persists in the presence of a scalar mass and a self-interaction,
we have computed the domains on four slices of the space of couplings, placed on both sides of the
tilt thresholds of Sec.~\ref{sec:wnl}: S1a ($\kappa_{\mu}=0$, $\kappa_{\lambda}=0.05$), for which
the tilt changes sign at $\beta=0.356$; S1b ($\kappa_{\mu}=0$, $\kappa_{\lambda}=0.10$), which is
supercritical for all $\beta\geq0$; S2 ($\kappa_{\mu}=6.25\times10^{-4}$, $\kappa_{\lambda}=0$);
and S3 ($\kappa_{\mu}=6.25\times10^{-4}$, $\kappa_{\lambda}=0.10$). The results are collected in
Table~\ref{tab:domains-massive}. In every case to which the tilt laws of Sec.~\ref{sec:wnl} apply,
they correctly predict the direction of the branch at the threshold, and at $\beta=0$ the two
completions again coincide.

\begin{table}[t]
\caption{Domains of existence $\mathcal{E}_{X}$ (in $\hM$) in the massive and self-interacting
sector for the quadratic truncation (Q) and the exponential completion (E). The fifth column
indicates whether the branch is subcritical or supercritical at the threshold, which lies at
$\hM_{c}=1.173944$ for $\kappa_{\mu}=0$ and at $\hM_{c}=1.149333$ for
$\kappa_{\mu}=6.25\times10^{-4}$.}
\label{tab:domains-massive}
\begin{ruledtabular}
\begin{tabular}{cccccccc}
Slice & $\kappa_{\mu}$ & $\kappa_{\lambda}$ & $\beta$ & Branch & $\mathcal{E}_{\mathrm{Q}}$ &
$\mathcal{E}_{\mathrm{E}}$ & Relation\\
\hline
S1a & $0$ & $0.05$ & $0$ & sub & $[1.173944,\,1.211540]$ & $[1.173944,\,1.211540]$ & $\mathrm{Q}=\mathrm{E}$\\
S1a & $0$ & $0.05$ & $0.3$ & sub & $[1.173944,\,1.206171]$ & $[1.173944,\,1.206214]$ & $\mathrm{E}\supsetneq\mathrm{Q}$\\
S1a & $0$ & $0.05$ & $1$ & super & $[1.147693,\,1.173944]$ & $[1.151715,\,1.173944]$ & $\mathrm{Q}\supsetneq\mathrm{E}$\\
S1b & $0$ & $0.10$ & $0$ & super & $[1.163251,\,1.174809]$ & $[1.163251,\,1.174809]$ & $\mathrm{Q}=\mathrm{E}$\\
S1b & $0$ & $0.10$ & $1$ & super & $[1.116578,\,1.173944]$ & $[1.121369,\,1.173944]$ & $\mathrm{Q}\supsetneq\mathrm{E}$\\
S1b & $0$ & $0.10$ & $2$ & super & $[0.954124,\,1.173944]$ & $[1.011333,\,1.173944]$ & $\mathrm{Q}\supsetneq\mathrm{E}$\\
S2 & $6.25\times10^{-4}$ & $0$ & $0$ & sub & $[1.149333,\,1.244530]$ & $[1.149333,\,1.244530]$ & $\mathrm{Q}=\mathrm{E}$\\
S2 & $6.25\times10^{-4}$ & $0$ & $1$ & sub & $[1.149333,\,1.201420]$ & $[1.149333,\,1.206662]$ & $\mathrm{E}\supsetneq\mathrm{Q}$\\
S2 & $6.25\times10^{-4}$ & $0$ & $2$ & super & $[0.973567,\,1.149333]$ & $[1.058475,\,1.149333]$ & $\mathrm{Q}\supsetneq\mathrm{E}$\\
S3 & $6.25\times10^{-4}$ & $0.10$ & $0$ & super & $[1.146798,\,1.163909]$ & $[1.146798,\,1.163909]$ & $\mathrm{Q}=\mathrm{E}$\\
S3 & $6.25\times10^{-4}$ & $0.10$ & $1$ & super & $[1.109932,\,1.149333]$ & $[1.113757,\,1.149333]$ & $\mathrm{Q}\supsetneq\mathrm{E}$\\
S3 & $6.25\times10^{-4}$ & $0.10$ & $2$ & super & $[0.954002,\,1.149333]$ & $[1.008423,\,1.149333]$ & $\mathrm{Q}\supsetneq\mathrm{E}$\\
\end{tabular}
\end{ruledtabular}
\end{table}

The pattern found for a massless scalar persists. On all six supercritical branches with
$\beta>0$ the domain of the quadratic truncation strictly contains that of the exponential
completion, with margins between $3.8\times10^{-3}$ and $8.5\times10^{-2}$ that grow with $\beta$.
The exponential completion has the larger domain only on the two subcritical branches with
$\beta>0$, namely S1a at $\beta=0.3$, close to the tilt threshold of that slice, where the margin is
$(4.507\pm0.018)\times10^{-5}$ as determined by a dense resolution of the endpoints, and S2 at
$\beta=1$, where it is $5.2\times10^{-3}$. These subcritical cases are the analogs of regime (iii)
of the massless sector.

In summary, the exponential completion has a larger domain of existence than the quadratic
truncation only on subcritical branches. On supercritical branches the quadratic truncation always
reaches lower masses, and its domain strictly contains that of the exponential completion, by
margins that grow with $\beta$, except in the narrow massless window $\btilt<\beta<\beta_{c2}$, in
which the exponential branch ends above the threshold and neither domain contains the other. In Sec.~\ref{sec:mechanism} we
explain this pattern and show that it is governed by the parameter $\gamma$, and in
Sec.~\ref{sec:stability} we show that the black holes that make up the advantages of the
exponential completion are radially unstable.

\section{Ordering mechanism and the completion law}
\label{sec:mechanism}
The comparisons of Sec.~\ref{sec:existence} can be traced to two elementary properties of the
branches, both of which originate in the Ricci sector. In this section we identify them, show that
they are controlled by the parameter $\gamma$ of Eq.~\eqref{eq:F-family}, and test the resulting
predictions with the completions R and M, which we constructed for this purpose.

\subsection{Shift and length effects}
\label{subsec:shift-length}

We compare two completions with the same couplings at the same horizon value of the scalar and
define $\Delta\hM_{X}(\phi_{h})\equiv\hM_{X}(\phi_{h})-\hM_{\mathrm{Q}}(\phi_{h})$. For the
exponential completion we find two effects in all 21 cases examined, namely 16 massless cases with
$0.5\leq\beta\leq3$ and five cases from the massive and self-interacting slices of
Sec.~\ref{subsec:massive-sector}:
\begin{enumerate}
\renewcommand{\labelenumi}{(\alph{enumi})}
\item \emph{Shift}: $\Delta\hM_{\mathrm{E}}(\phi_{h})>0$ on the entire common range of $\phi_{h}$,
and at small amplitude $\Delta\hM_{\mathrm{E}}\simeq c_{4}(\beta)\,\phi_{h}^{4}$ with $c_{4}>0$.
\item \emph{Length}: the branch of the exponential completion terminates at a smaller value of
$\phi_{h}$ than that of the quadratic truncation.
\end{enumerate}
These two properties determine the ordering of the domains. By (a), the minimum of
$\hM_{\mathrm{E}}$ over its branch exceeds the minimum of $\hM_{\mathrm{Q}}$ over the same range of
$\phi_{h}$, and by (b) the latter cannot be smaller than the minimum of $\hM_{\mathrm{Q}}$ over its
longer branch. Hence the quadratic truncation always reaches lower masses, which is the origin of
the strict containments on supercritical branches found in Sec.~\ref{sec:existence}. At the upper
edge of subcritical branches the two effects compete: the upward shift favors the exponential
completion, whereas its shorter branch favors the quadratic truncation. As $\beta$ approaches
$\btilt$ the subcritical branches flatten, so that less mass is forgone by terminating earlier while
the shift keeps growing; this is consistent with the crossover at $\beta_{\times}\simeq0.63$. In the
window $\btilt<\beta<\beta_{c2}$, finally, the upward-shifted branch of the exponential completion
ends above the threshold, and neither domain contains the other.

Both effects are properties of the Ricci sector. The potential affects the domains only through the
direction of the bifurcation, and for the couplings considered the shift is only weakly sensitive to
the mass and to the self-interaction: at $\beta=1$ and $2$ the ranges of
$(\hM_{\mathrm{E}}-\hM_{\mathrm{Q}})/\phi_{h}^{4}$ over $0.15\leq\phi_{h}\leq0.5$ on the massive
and self-interacting slices agree with the massless ones to within about $10\%$.

\subsection{The completion law}
\label{subsec:completion-law}

Both effects are controlled by $\gamma$. Since $\delta F\equiv F_{X}-F_{\mathrm{Q}}=\gamma\beta^{2}
\phi^{4}+O(\phi^{6})$ enters the metric equations at fourth order and the scalar equation at fifth
order in the amplitude (Sec.~\ref{subsec:completions}), the leading change of any quantity along the
branch is linear in $\delta F$ and hence in $\gamma$. For the mass this gives the completion law
\begin{equation}
\hM_{X}(\phi_{h})-\hM_{\mathrm{Q}}(\phi_{h})=32\gamma\,c_{4}(\beta)\,\phi_{h}^{4}+O(\phi_{h}^{6}),
\label{eq:completion-law}
\end{equation}
in which $c_{4}(\beta)$ is the coefficient of the exponential completion ($\gamma=1/32$). The
linearity in $\gamma$ is exact at this order, whereas $c_{4}(\beta)$ must be computed; it is positive
in all cases that we have considered. Expanding the regularity function~\eqref{eq:S-einstein} at
equal $\phi_{h}$ and $\zeta$ similarly gives
\begin{equation}
\mathcal{S}_{h}^{X}-\mathcal{S}_{h}^{\mathrm{Q}}=\gamma\beta^{2}\,(1-12\beta-48\zeta)\,\phi_{h}^{4}
r_{h}^{4}+O(\phi_{h}^{6}),
\label{eq:length-law}
\end{equation}
which holds for arbitrary completions. Along all branches $\zeta\gtrsim0.17$, so the bracket is
negative: a completion with $\gamma>0$ reaches the regularity bound earlier, and one with
$\gamma<0$ later. Both effects are therefore monotonic in $\gamma$, and increasing $\gamma$ shifts
the branch to larger masses and shortens it. In physical terms, a completion with $\gamma>0$
saturates the enhancement of the effective gravitational coupling $G_{\mathrm{eff}}=G/F$ by the
scalar: the black holes are heavier at given $\phi_{h}$, and the Ricci contributions that relax the
regularity bound (Sec.~\ref{subsec:regularity}) are reduced.

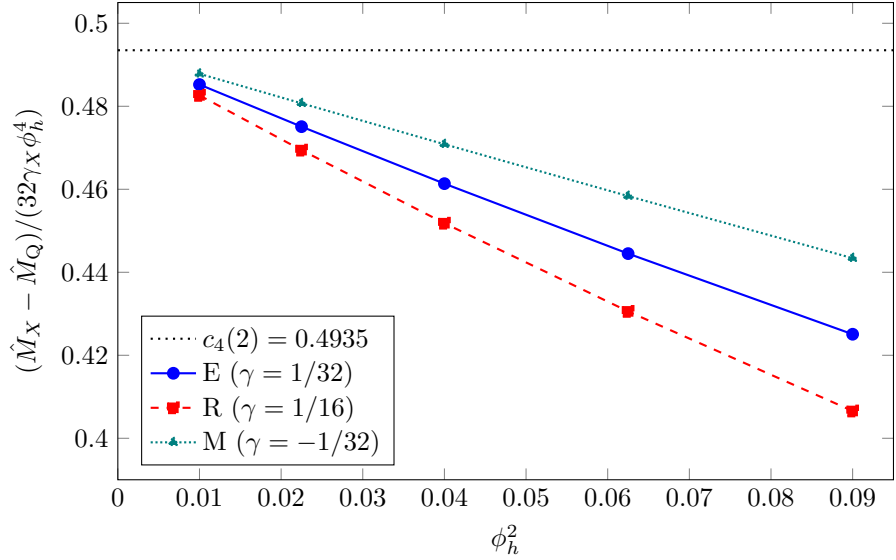
\begin{figure}[t]
\centering
\begin{tikzpicture}
\begin{axis}[width=0.72\textwidth, height=0.48\textwidth,
  xmin=0, xmax=0.095, ymin=0.39, ymax=0.505,
  xlabel={$\phi_{h}^{2}$}, ylabel={$(\hM_{X}-\hM_{\mathrm{Q}})/(32\gamma_{X}\phi_{h}^{4})$},
  legend pos=south west, legend cell align=left, tick label style={font=\small},
  scaled x ticks=false, xticklabel style={/pgf/number format/fixed}]
\addplot[black, dotted, thick] coordinates {(0,0.4935) (0.095,0.4935)};
\addlegendentry{$c_{4}(2)=0.4935$}
\addplot[blue, thick, mark=*] coordinates {(0.0100,0.485234) (0.0225,0.475087) (0.0400,0.461355) (0.0625,0.444497) (0.0900,0.425067)};
\addlegendentry{E ($\gamma=1/32$)}
\addplot[red, thick, dashed, mark=square*] coordinates {(0.0100,0.482623) (0.0225,0.469434) (0.0400,0.451823) (0.0625,0.430574) (0.0900,0.406574)};
\addlegendentry{R ($\gamma=1/16$)}
\addplot[teal, thick, densely dotted, mark=triangle*] coordinates {(0.0100,0.487826) (0.0225,0.480712) (0.0400,0.470851) (0.0625,0.458358) (0.0900,0.443396)};
\addlegendentry{M ($\gamma=-1/32$)}
\end{axis}
\end{tikzpicture}
\caption{Test of the completion law~\eqref{eq:completion-law} at $\beta=2$. The mass shift of each
completion relative to the quadratic truncation, divided by $32\gamma_{X}\phi_{h}^{4}$, is shown for
the exponential (E), rational (R), and negative-$\gamma$ (M) completions. The three curves differ at
finite amplitude by terms of order $\phi_{h}^{2}$ but extrapolate to the common value
$c_{4}(2)=0.4935\pm0.0001$ (dotted line), as predicted: the ratios of the extrapolated coefficients
agree with the predicted values $+2$ and $-1$ to within $3\times10^{-4}$.}
\label{fig:collapse}
\end{figure}

The completions R ($\gamma=1/16$) and M ($\gamma=-1/32$) provide a predictive test of
Eq.~\eqref{eq:completion-law}, whose coefficients must satisfy $c_{4}^{\mathrm{R}}/c_{4}^{\mathrm{E}}=2$
and $c_{4}^{\mathrm{M}}/c_{4}^{\mathrm{E}}=-1$. Figure~\ref{fig:collapse} shows the mass shifts at
$\beta=2$, each divided by $32\gamma_{X}\phi_{h}^{4}$. The three curves differ at finite amplitude
by terms of order $\phi_{h}^{2}$, as expected, but extrapolate to the common value
$c_{4}(2)=0.4935\pm0.0001$, and the ratios agree with the predicted values to within
$3\times10^{-4}$.

\subsection{Consequences for the domains}
\label{subsec:gamma-ordering}

\begin{figure}[t]
\centering
\begin{tikzpicture}
\begin{axis}[width=0.78\textwidth, height=0.52\textwidth,
  xmin=0, xmax=1.2, ymin=0.78, ymax=1.19,
  xlabel={$\phi_{h}$}, ylabel={$\hM=M/\sqrt{\alpha}$},
  legend pos=south west, legend cell align=left, tick label style={font=\small}]
\addplot[black, dotted, thick] coordinates {(0,1.173944) (1.2,1.173944)};
\addlegendentry{threshold $\hM_{c}$}
\addplot[black, thick] coordinates {(0.0200,1.17372) (0.0419,1.17275) (0.0638,1.17161) (0.0857,1.16969) (0.1076,1.16745) (0.1295,1.16462) (0.1514,1.16133) (0.1733,1.15764) (0.1952,1.15341) (0.2171,1.14893) (0.2390,1.14386) (0.2609,1.13866) (0.2829,1.13290) (0.3048,1.12703) (0.3267,1.12078) (0.3486,1.11438) (0.3705,1.10776) (0.3924,1.10099) (0.4143,1.09412) (0.4362,1.08715) (0.4581,1.08014) (0.4800,1.07312) (0.5019,1.06611) (0.5238,1.05919) (0.5457,1.05232) (0.5676,1.04562) (0.5895,1.03905) (0.6114,1.03265) (0.6333,1.02651) (0.6552,1.02053) (0.6771,1.01493) (0.6990,1.00945) (0.7209,1.00451) (0.7428,0.99968) (0.7648,0.99540) (0.7867,0.99133) (0.8086,0.98774) (0.8305,0.98450) (0.8524,0.98164) (0.8743,0.97928) (0.8962,0.97717) (0.9181,0.97570) (0.9400,0.97436) (0.9619,0.97379) (0.9835,0.97331) (0.9837,0.97331) (0.9838,0.97331) (0.9838,0.97331) (0.9838,0.97331)};
\addlegendentry{Q ($\gamma=0$)}
\addplot[blue, thick, dashed] coordinates {(0.0200,1.17372) (0.0380,1.17293) (0.0560,1.17213) (0.0740,1.17074) (0.0920,1.16918) (0.1100,1.16722) (0.1280,1.16495) (0.1460,1.16247) (0.1641,1.15958) (0.1821,1.15662) (0.2001,1.15320) (0.2181,1.14978) (0.2361,1.14600) (0.2541,1.14218) (0.2721,1.13818) (0.2901,1.13408) (0.3081,1.12992) (0.3261,1.12569) (0.3441,1.12146) (0.3621,1.11722) (0.3801,1.11299) (0.3981,1.10888) (0.4161,1.10477) (0.4341,1.10086) (0.4522,1.09700) (0.4702,1.09335) (0.4882,1.08987) (0.5062,1.08654) (0.5242,1.08352) (0.5422,1.08056) (0.5602,1.07810) (0.5782,1.07564) (0.5962,1.07374) (0.6142,1.07191) (0.6322,1.07056) (0.6502,1.06943) (0.6682,1.06864) (0.6862,1.06825) (0.7000,1.06795) (0.7042,1.06804) (0.7222,1.06843) (0.7403,1.06883) (0.7583,1.07000) (0.7763,1.07116) (0.7943,1.07275) (0.8122,1.07452) (0.8123,1.07453) (0.8123,1.07453) (0.8123,1.07453)};
\addlegendentry{E ($\gamma=1/32$)}
\addplot[red, thick, dashdotted] coordinates {(0.0200,1.17372) (0.0365,1.17300) (0.0530,1.17227) (0.0695,1.17114) (0.0860,1.16973) (0.1026,1.16822) (0.1191,1.16619) (0.1356,1.16417) (0.1521,1.16177) (0.1686,1.15923) (0.1851,1.15658) (0.2016,1.15367) (0.2181,1.15076) (0.2347,1.14764) (0.2512,1.14449) (0.2677,1.14130) (0.2842,1.13807) (0.3007,1.13484) (0.3172,1.13167) (0.3337,1.12849) (0.3502,1.12545) (0.3667,1.12247) (0.3833,1.11957) (0.3998,1.11692) (0.4163,1.11427) (0.4328,1.11197) (0.4493,1.10976) (0.4658,1.10773) (0.4823,1.10605) (0.4988,1.10437) (0.5154,1.10325) (0.5319,1.10217) (0.5484,1.10142) (0.5649,1.10100) (0.5800,1.10062) (0.5814,1.10064) (0.5979,1.10092) (0.6144,1.10120) (0.6309,1.10195) (0.6475,1.10295) (0.6640,1.10412) (0.6805,1.10583) (0.6970,1.10753) (0.7135,1.10976) (0.7300,1.11210) (0.7463,1.11450) (0.7464,1.11452) (0.7465,1.11453) (0.7465,1.11454)};
\addlegendentry{R ($\gamma=1/16$)}
\addplot[teal, very thick, densely dotted] coordinates {(0.0200,1.17372) (0.0463,1.17255) (0.0726,1.17082) (0.0989,1.16849) (0.1252,1.16504) (0.1515,1.16104) (0.1778,1.15640) (0.2041,1.15073) (0.2304,1.14453) (0.2568,1.13768) (0.2831,1.12989) (0.3094,1.12162) (0.3357,1.11270) (0.3620,1.10299) (0.3883,1.09285) (0.4146,1.08212) (0.4409,1.07078) (0.4672,1.05911) (0.4935,1.04696) (0.5198,1.03442) (0.5461,1.02166) (0.5724,1.00862) (0.5987,0.99543) (0.6250,0.98217) (0.6513,0.96888) (0.6776,0.95569) (0.7040,0.94259) (0.7303,0.92978) (0.7566,0.91730) (0.7829,0.90508) (0.8092,0.89348) (0.8355,0.88237) (0.8618,0.87167) (0.8881,0.86189) (0.9144,0.85268) (0.9407,0.84399) (0.9670,0.83639) (0.9933,0.82936) (1.0196,0.82288) (1.0459,0.81726) (1.0722,0.81231) (1.0985,0.80794) (1.1248,0.80407) (1.1350,0.80265) (1.1512,0.80054) (1.1550,0.80004) (1.1750,0.79746) (1.1775,0.79712)};
\addlegendentry{M ($\gamma=-1/32$)}
\end{axis}
\end{tikzpicture}
\caption{Fundamental branches of the four completions for a massless scalar at $\beta=2$: normalized
mass versus the horizon value of the scalar. All branches leave the threshold (dotted line) in the
supercritical direction. At equal $\phi_{h}$, completions with larger $\gamma$ are heavier (shift
effect), and their branches are shorter (length effect); both effects raise the minimum of $\hM$.
The branches of Q and E end by discriminant closure; the branch of R is shown beyond its turning
point, and that of M up to the last computed solution, where $F(\phi_{h})=0.07$.}
\label{fig:branches}
\end{figure}
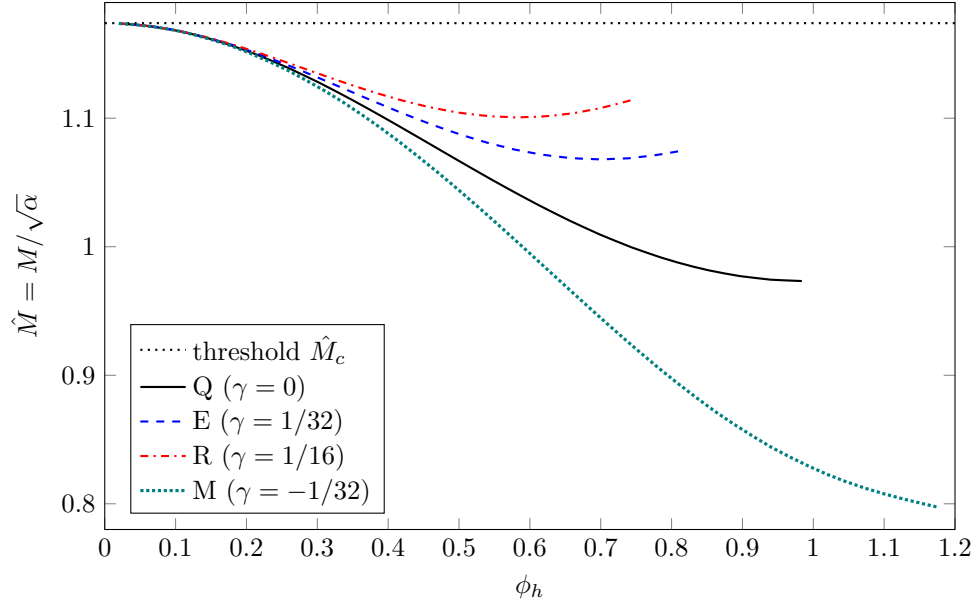

Because both effects are monotonic in $\gamma$, the lower edges of the domains of existence on
supercritical branches are ordered by $\gamma$. Figure~\ref{fig:branches} displays the four
branches at $\beta=2$, and Table~\ref{tab:gamma-order} lists the lower edges at $\beta=1.5$, $2$, and
$3$; in every case the lower edge increases monotonically with $\gamma$. The branches of R were
followed beyond their turning points, which set their lower edges, but not to their ends. The
branches of M behave differently from all others: they do not terminate by discriminant closure but
continue to large horizon values of the scalar, with $\mathcal{S}_{h}/\mathcal{Q}_{h}$ of order
unity, while $F(\phi_{h})$ decreases to $0.15$ ($\beta=1.5$) and $0.07$ ($\beta=2$ and $3$) on the
last solutions that we could compute. Along these solutions $|\phi|$ decreases monotonically
outward, so $F>0$ everywhere. The data indicate that the branches approach the zero of $F$, where the
effective gravitational coupling diverges on the horizon; we have not resolved this final approach,
so for M the entries of Table~\ref{tab:gamma-order} are upper bounds on the lower edge. These bounds
suffice for the ordering, which they satisfy with wide margins.

\begin{table}[t]
\caption{Lower edges of the domains of existence, in $\hM=M/\sqrt{\alpha}$, for the four
completions and a massless scalar. The lower edges of Q, E, and R are minima of $\hM$ along the
branch; for M, whose branch could not be followed to its end (see text), the entries are upper
bounds given by the last computed solution. In Sec.~\ref{sec:stability} we show that these are also
the lower edges of the domains of radially stable black holes.}
\label{tab:gamma-order}
\begin{ruledtabular}
\begin{tabular}{ccccc}
$\beta$ & M ($\gamma=-1/32$) & Q ($\gamma=0$) & E ($\gamma=1/32$) & R ($\gamma=1/16$)\\
\hline
$1.5$ & $\leq0.951811$ & $1.103793$ & $1.141222$ & $1.152329$\\
$2$ & $\leq0.797123$ & $0.973308$ & $1.067953$ & $1.100608$\\
$3$ & $\leq0.629384$ & $0.786334$ & $0.935550$ & $0.998529$\\
\end{tabular}
\end{ruledtabular}
\end{table}

The completion law thus reduces the dependence of the domains of existence on the nonlinear
completion to a single number, $\gamma$, at the order at which the completions first differ. Its
consequences for radial stability are the subject of the next section.

\section{Radial stability and hyperbolicity}
\label{sec:stability}
We now determine which of the black holes constructed in Secs.~\ref{sec:existence}
and~\ref{sec:mechanism} are stable against spherically symmetric perturbations. We derive a master
equation for such perturbations that holds for arbitrary couplings, count its unstable modes in two
independent ways, and determine the domains of radially stable black holes. Black-hole
perturbations in the Horndeski class have been studied in general in
Refs.~\cite{KMS2012,KMS2014}, and radial and angular stability conditions for scalarized black
holes with a quadratic Ricci coupling have been obtained in Ref.~\cite{Minamitsuji2024}.

\subsection{Master equation for radial perturbations}
\label{subsec:master}

We write time-dependent, spherically symmetric configurations in the areal gauge,
\begin{equation}
ds^{2}=-A(t,r)\,dt^{2}+\frac{dr^{2}}{B(t,r)}+r^{2}d\Omega^{2},\qquad \phi=\phi(t,r),
\label{eq:td-ansatz}
\end{equation}
in which $\delta g_{tr}$ can be set to zero for $\ell=0$ perturbations, and perturb a static
solution $(A_{0},B_{0},\phi_{0})$, with $A_{0}=e^{2\delta}N$ and $B_{0}=N$ in the notation of
Sec.~\ref{sec:static}, as $A=A_{0}(1+\varepsilon a)$, $B=B_{0}(1+\varepsilon b)$, and
$\phi=\phi_{0}+\varepsilon q$. At linear order, the $tr$ component of the metric equations is an
exact time derivative, $E_{tr}^{(1)}=\partial_{t}(c_{1}b+c_{2}q+c_{3}q')$, in which $a$ does not
appear; for perturbations with nontrivial time dependence it determines $b$ algebraically in terms
of $q$ and $q'$. The $rr$ component contains $a$ only through $a'$, which it determines
algebraically in terms of $b$, $q$, $q'$, and $\partial_{t}^{2}q$. Substituting into the linearized
scalar equation yields the master equation
\begin{equation}
K_{tt}\,\partial_{t}^{2}q+K_{rr}\,\partial_{r}^{2}q+K_{r}\,\partial_{r}q+K_{0}\,q=0 .
\label{eq:master}
\end{equation}
Its coefficients depend on the background and on $F$, $f$, $V$, and their derivatives up to third
order. The function $f$ enters only through its derivatives, reflecting the invariance of the action
under $f\to f+\mathrm{const}$, since $\sqrt{-g}\,\GB$ is a total derivative. The explicit coefficients
are given in Appendix~\ref{app:master}.

We have verified Eq.~\eqref{eq:master} in four independent ways. (i) No mixed derivative
$\partial_{t}^{2}\partial_{r}q$ appears; its coefficient vanishes to a relative accuracy of
$10^{-17}$, so the equation is of second order, as required for a Horndeski theory. (ii) The
time-dependent field equations from which it is derived satisfy the Noether
identity~\eqref{eq:noether} off shell, to $10^{-13}$. (iii) For $F=1$ and $f=V=0$ it reduces
exactly to the $\ell=0$ Klein--Gordon equation on the Schwarzschild background. (iv) For arbitrary
couplings on the Schwarzschild background, where $\phi_{0}=0$ and the metric perturbations decouple,
it reduces to Eq.~\eqref{eq:linear}, in which the Ricci coupling is absent because $R=0$.

The principal part of Eq.~\eqref{eq:master} defines the radial characteristic speed
$c^{2}\equiv-K_{rr}/K_{tt}$, and the equation is hyperbolic where $c^{2}>0$. In general relativity
$c^{2}=A_{0}B_{0}$, and we use the normalized speed $s\equiv c^{2}/(A_{0}B_{0})$, which equals unity
in the absence of couplings.

Unstable modes have the form $q=e^{\Omega t}u(r)$ with $\Omega>0$. Regularity on the future horizon
requires $u\propto(r-r_{h})^{\Omega/(2\kappa)}$, where $\kappa=2\pi T$ is the surface gravity, and
at infinity $u\propto e^{-kr}/r$ with $k=(\Omega^{2}+\mu^{2})^{1/2}$. We count unstable modes in two
independent ways: by scanning $\Omega$ for the zeros of a growth indicator built from the two
boundary conditions, and, using the Sturm oscillation theorem, by counting the nodes of the static
($\Omega=0$) solution that is regular on the horizon. The two counts agree in all cases, provided the
radial domain is large enough to resolve very slowly growing modes. For example, on the branch of
the quadratic truncation with $\beta=1.2$ a mode is born at the turning point: at $\phi_{h}=0.55$,
just beyond it, the growth rate is $\Omega M=3.2\times10^{-4}$, and the corresponding static solution
has its node at $r\approx124\,r_{h}$.

\subsection{Exchange of stability}
\label{subsec:exchange}

Near the threshold, the direction of the bifurcation (Sec.~\ref{sec:wnl}) decides stability:
subcritical branches carry one unstable mode, whereas supercritical branches are stable. For the
quadratic truncation at $\phi_{h}=0.2$ we find one unstable mode for $\beta=1.05$, $1.10$, and
$1.11$, and none for $\beta=1.13$ (and none for $\beta=1.12$ before its turning point). The change
of stability therefore lies between $\beta=1.11$ and $1.12$, in agreement with
$\btilt=1.1156$ of Eq.~\eqref{eq:beta-tilt}. Close to $\btilt$ the growth rate becomes very small,
$\Omega r_{h}=5.5\times10^{-5}$ at $\beta=1.10$, and resolving it requires a radial domain extending
to $1.8\times10^{5}\,r_{h}$.

Along supercritical branches, stability is lost exactly at the first minimum of $\hM$: the number
of unstable modes changes from zero to one across the turning point, as expected from the
turning-point criterion~\cite{Sorkin1981}. We have verified this at all 11 turning points examined:
for the quadratic truncation at $\beta=1.12$ (turning point at $\phi_{h}=0.1285$,
$\hM=1.1739275$) and $\beta=1.13$ ($\phi_{h}=0.2310$, $\hM=1.1737652$), for both completions at
$\beta=1.2$, for the exponential completion at $\beta=2$, on the massive and self-interacting
slices S1b and S3, and for the completion R at $\beta=1.5$, $2$, and $3$. Subcritical branches, by
contrast, are unstable at all sampled points, from $15\%$ to $98\%$ of their length. The radially
stable part of a supercritical branch is therefore the segment between the threshold and the first
minimum of $\hM$, or the end of the branch if $\hM$ decreases monotonically up to it, and the domain
of radially stable black holes is $\mathcal{D}_{X}=[\hM_{\min},\hM_{c})$.

These findings are consistent with previous work. The instability of the part of the branch beyond
the mass minimum was reported for a quadratic Ricci coupling in Ref.~\cite{Antoniou2022}, and the
marginal stability at the turning point in the presence of self-interactions in
Ref.~\cite{Macedo2019}. For $\beta=1.2$ our turning point at $\hM=1.168503$ agrees with the value
$\hM\approx1.168$ reported in Ref.~\cite{Antoniou2022}. The new elements are the exact threshold of
stability at the bifurcation, Eq.~\eqref{eq:beta-tilt}, and the verification of the turning-point
rule across completions and couplings.

\subsection{Radially stable domains}
\label{subsec:stable-domains}

Table~\ref{tab:stable} lists the radially stable domains for the cases compared in
Sec.~\ref{sec:existence}. In every case the radially stable domain of the quadratic truncation
contains that of the exponential completion, strictly whenever it is nonempty. The advantages of the
exponential completion in the domains of existence consist entirely of radially unstable black
holes: the upper edges of subcritical branches [regime (iii) of Sec.~\ref{subsec:massless-domains}
and the subcritical cases of Table~\ref{tab:domains-massive}] are unstable because the whole branch
is, and in the window $\btilt<\beta<\beta_{c2}$ the part of the exponential branch above the
threshold lies beyond its turning point. At $\beta=1.2$, for instance, the domains of existence are
not nested, whereas the radially stable domains satisfy
$\mathcal{D}_{\mathrm{Q}}\supsetneq\mathcal{D}_{\mathrm{E}}$.

\begin{table}[t]
\caption{Radially stable domains $\mathcal{D}_{X}$ (in $\hM=M/\sqrt{\alpha}$) of the quadratic
truncation (Q) and the exponential completion (E). The upper edge is the threshold $\hM_{c}$ and the
lower edge the first minimum of $\hM$ or the end of the branch. Subcritical branches are unstable
throughout, so that their stable domains are empty.}
\label{tab:stable}
\begin{ruledtabular}
\begin{tabular}{ccccc}
Case & $\beta$ & $\mathcal{D}_{\mathrm{Q}}$ & $\mathcal{D}_{\mathrm{E}}$ & Relation\\
\hline
massless & $1.0$ & $\emptyset$ & $\emptyset$ & equal\\
S2 ($\kappa_{\mu}=6.25\times10^{-4}$) & $1.0$ & $\emptyset$ & $\emptyset$ & equal\\
massless & $1.2$ & $[1.168503,\,1.173944)$ & $[1.171644,\,1.173944)$ &
$\mathcal{D}_{\mathrm{Q}}\supsetneq\mathcal{D}_{\mathrm{E}}$\\
massless & $2.0$ & $[0.973308,\,1.173944)$ & $[1.067953,\,1.173944)$ &
$\mathcal{D}_{\mathrm{Q}}\supsetneq\mathcal{D}_{\mathrm{E}}$\\
S1b ($\kappa_{\lambda}=0.10$) & $1.0$ & $[1.116578,\,1.173944)$ & $[1.121369,\,1.173944)$ &
$\mathcal{D}_{\mathrm{Q}}\supsetneq\mathcal{D}_{\mathrm{E}}$\\
S3 ($\kappa_{\mu}=6.25\times10^{-4}$, $\kappa_{\lambda}=0.10$) & $2.0$ & $[0.954002,\,1.149333)$ &
$[1.008423,\,1.149333)$ & $\mathcal{D}_{\mathrm{Q}}\supsetneq\mathcal{D}_{\mathrm{E}}$\\
\end{tabular}
\end{ruledtabular}
\end{table}

The same rule applies to the completions of Sec.~\ref{sec:mechanism}. For Q, E, and R the lower
edges listed in Table~\ref{tab:gamma-order} are the points at which stability is lost, so they are
also the lower edges of the radially stable domains. For M, every solution computed along the
continued branches is radially stable: no unstable mode appears at any of the 23 sampled solutions
at $\beta=1.5$, $2$, and $3$, up to the last computed one, so its radially stable domain extends at
least to the bounds of Table~\ref{tab:gamma-order}. Consequently, at $\beta=1.5$, $2$, and $3$,
\begin{equation}
\mathcal{D}_{\mathrm{M}}\supsetneq\mathcal{D}_{\mathrm{Q}}\supsetneq\mathcal{D}_{\mathrm{E}}
\supsetneq\mathcal{D}_{\mathrm{R}},
\label{eq:gamma-ordering}
\end{equation}
that is, the domain of radially stable black holes shrinks monotonically as $\gamma$ increases,
and the exponential completion supports a smaller domain of radially stable black holes than the
quadratic truncation.

For $\kappa_{\mu}=6.25\times10^{-4}$ and $\beta=0$, that is, $\hat{\mu}=0.05$ in the notation of
Ref.~\cite{Macedo2019}, the tilt law~\eqref{eq:massive-law} predicts that the branch becomes
supercritical, and hence radially stable near the threshold, for $\kappa_{\lambda}>0.078739$, that
is, $\hat{\lambda}>0.315$. Counting unstable modes at $\phi_{h}=0.10$, we find one unstable mode for
$\hat{\lambda}=0.28$ and $0.30$ and none for $\hat{\lambda}=0.33$ and $0.36$. The change of stability
therefore lies between $\hat{\lambda}=0.30$ and $0.33$, bracketing the prediction, which is thus
confirmed by a method independent of the fits that determine Eq.~\eqref{eq:massive-law}.

\subsection{Hyperbolicity near the ends of the branches}
\label{subsec:hyperbolicity}

All sampled backgrounds are hyperbolic, $c^{2}>0$ at all radii, including those close to the ends of
the branches. As a branch approaches discriminant closure, however, the normalized speed $s$
develops a minimum in a thin layer adjacent to the horizon, which deepens as the closure is
approached. For $\beta=0$, $\min s=0.942$, $0.838$, $0.737$, and $0.648$ as the normalized
discriminant $D_{2}/B_{2}^{2}=\mathcal{S}_{h}/(r_{h}^{2}\mathcal{Q}_{h})$ decreases through $0.24$,
$0.063$, $0.016$, and $0.0034$, while the location of the minimum moves from $r-r_{h}=0.024\,r_{h}$
to $0.0012\,r_{h}$. For the quadratic truncation at $\beta=1.2$ we find $\min s=0.961$, $0.872$, and
$0.769$ at $\mathcal{S}_{h}/(r_{h}^{2}\mathcal{Q}_{h})=0.29$, $0.082$, and $0.022$. The decrease is
roughly logarithmic in the discriminant, so that radial hyperbolicity is lost at, or immediately
before, discriminant closure; numerically, the two cannot be separated. This agrees with the loss
of hyperbolicity in a region just outside the horizon found in
Refs.~\cite{BlazquezSalcedo2018,Rossi2026}, and with the loss of hyperbolicity observed in
dynamical evolutions~\cite{RipleyPretorius2019a,RipleyPretorius2019b,EastRipley2021}.

These observations do not affect our conclusions. Edges set by turning points lie far from closure.
An edge set by the end of a branch, such as that of the quadratic truncation at $\beta=2$, would
move by at most about $10^{-4}$ in $\hM$ even if hyperbolicity were lost where
$\mathcal{S}_{h}/(r_{h}^{2}\mathcal{Q}_{h})\approx10^{-3}$, far below the corresponding margins,
which exceed $5\times10^{-2}$. Along the continued branches of M, which do not approach discriminant
closure, $s$ remains above $0.92$. Stability against non-radial perturbations is discussed in
Sec.~\ref{sec:discussion}.

\section{Discussion and conclusion}
\label{sec:discussion}
In this section we compare our results with previous work, discuss their frame independence and
their interpretation from the viewpoint of effective field theory, state their limitations, and
summarize our conclusions.

\subsection{Comparison with previous results}
\label{subsec:benchmarks}

Where our results overlap with previous work, they reproduce the available benchmarks. The
threshold~\eqref{eq:threshold} corresponds to $\eta/M^{2}=2.9025$ in the normalization of
Ref.~\cite{Silva2018}, in agreement with the value $2.902$ quoted there, and the general
regularity condition of Sec.~\ref{subsec:regularity} reduces to the conditions of
Refs.~\cite{DonevaYazadjiev2018,AntoniouBakopoulosKanti2018,Silva2018,Antoniou2021,Macedo2019}
in the appropriate limits. For the pure scalar--Gauss--Bonnet theory the fundamental branch ends at
$M/\lambda_{\mathrm{B}}=0.6368$, in agreement with the value $0.636$ of
Ref.~\cite{BlazquezSalcedo2018}, whereas the lower edge $\eta/M^{2}\approx2.53$ quoted in
Ref.~\cite{Silva2018} lies before the point at which the regularity bound is saturated
(Sec.~\ref{subsec:termination}). For a quadratic Ricci coupling, the exact value
$\btilt=1.1156$ refines the estimates $\beta\approx1.22$~\cite{Antoniou2021} and
$\beta\approx1.15$--$1.2$~\cite{Antoniou2022} obtained at finite amplitude, and our turning point at
$\hM=1.168503$ for $\beta=1.2$ agrees with Ref.~\cite{Antoniou2022}. The instability beyond the
mass minimum~\cite{Antoniou2022,Macedo2019} is recovered and verified across completions and
couplings (Sec.~\ref{subsec:exchange}).

Macedo \textit{et al.}~\cite{Macedo2019} showed that a quartic self-interaction can render
scalarized black holes stable. Our tilt laws quantify this effect. In their notation,
$\hat{\lambda}=4\kappa_{\lambda}$, the bifurcation becomes supercritical for
$\hat{\lambda}>4\kappa_{\lambda}^{*}$, which equals $0.2359$ for a massless scalar without Ricci
coupling [Eq.~\eqref{eq:kappa-star}] and $0.315$ for $\hat{\mu}=0.05$
[Eq.~\eqref{eq:massive-law}], the latter confirmed independently by counting unstable modes
(Sec.~\ref{subsec:stable-domains}). A comparison at the level of individual branches would
require the numerical data underlying the figures of Ref.~\cite{Macedo2019}, which are not
available to us.

The tilt laws also connect with the description of scalarization as a phase
transition~\cite{Herdeiro2026,Muniz2025,Unluturk2025,Ozinan2026}. In that language, a
supercritical bifurcation corresponds to a continuous transition, whereas a subcritical one
corresponds to a discontinuous transition, with stable Schwarzschild black holes coexisting with
unstable scalarized ones near the threshold. Equations~\eqref{eq:tilt-law},
\eqref{eq:quartic-law}, and~\eqref{eq:massive-law} thus determine the order of the transition for
black holes with a Ricci coupling and a self-interaction, complementing the numerical analysis of
Ref.~\cite{Herdeiro2026} for black holes without these couplings and the first-principles analysis
of Ref.~\cite{Ozinan2026} for neutron stars.

\subsection{Frame independence and the effective-field-theory perspective}
\label{subsec:eft}

The quantities that we compare, namely the horizon value of the scalar, the ADM mass, the
temperature, and the frequencies of perturbations, are the same in the Jordan and Einstein frames
(Sec.~\ref{subsec:einstein}); our conclusions therefore do not depend on the frame in which the
theory is formulated. In the Einstein frame the completions differ through the kinetic function
$K(\phi)$ at fourth order, Eq.~\eqref{eq:KX}, and in the Jordan frame through the saturation of the
effective gravitational coupling $G/F$; $\gamma$ is the common coefficient of both effects.

From the viewpoint of effective field theory, $\gamma$ is a coefficient of the same order in the
scalar amplitude as the nonlinear effects of the quadratic coupling itself. Our results show that
it cannot be set to zero without consequence: at fixed $\beta$, the domain of radially stable black
holes depends monotonically on $\gamma$, so the quadratic truncation ($\gamma=0$) is not a neutral
choice. Conclusions about stability derived with a quadratic Ricci coupling should therefore be
understood as conclusions about completions with $\gamma=0$ at this order; a completion with
$\gamma>0$, such as the exponential one, supports fewer radially stable black holes.

The same viewpoint limits the physical significance of the strongly coupled parts of the branches.
A natural measure of the curvature on the horizon in units of the Gauss--Bonnet length scale is
$\alpha K_{h}^{1/2}$, where $K_{h}$ is the Kretschmann scalar on the horizon. It equals $0.63$ at the
threshold, so scalarization is intrinsically a phenomenon at the scale set by $\alpha$. At the lower
edges of the radially stable domains, $\alpha K_{h}^{1/2}$ remains of order unity for the exponential
completion, between $0.7$ and $3.0$ for $1.2\leq\beta\leq3$, but grows rapidly for the quadratic
truncation, reaching $4.9$ at $\beta=1.5$, $35$ at $\beta=2$, and $1.4\times10^{2}$ at $\beta=3$.
The part of the radially stable domain of the quadratic truncation that lies beyond that of the
exponential completion therefore extends into a regime of increasingly strong curvature, in which
operators not contained in the action~\eqref{eq:action} could become important. This caveat does
not affect the exact bifurcation laws or the completion law, which are derived at small amplitude,
but it qualifies the physical interpretation of the lower edges at large $\beta$.

\subsection{Non-radial perturbations and other limitations}
\label{subsec:limitations}

The principal limitation of this work is that we have analyzed radial perturbations only. For the
quadratic truncation, radially stable scalarized black holes are known to be unstable against
non-radial perturbations in part of their domain. Quadrupolar ($\ell=2$) zero modes appear for
$\beta\geq1.3458$ and lie on the radially stable part of the branch for
$\beta\gtrsim1.52$~\cite{Kleihaus2023}, and hexadecapolar ($\ell=4$) instabilities have also been
found~\cite{BlazquezSalcedo2024GRG}. For $\beta=2$, the figures of
Refs.~\cite{Kleihaus2023,BlazquezSalcedo2024GRG} place the onset of the $\ell=2$ and $\ell=4$
instabilities at $M/\sqrt{\alpha}\approx0.98$ and $\approx1.01$, respectively, both within the
radially stable domain $[0.9733,\,1.1739)$ of Table~\ref{tab:stable}. Analyses of high multipoles
indicate a hierarchy of angular instabilities~\cite{Minamitsuji2024,BlazquezSalcedo2026hier}, and it
has been reported that along the branch stability is first lost to perturbations of high multipole
order~\cite{BlazquezSalcedo2026rev}. The domain of fully stable black holes of the quadratic
truncation is therefore smaller than its radially stable domain. For non-quadratic completions,
non-radial stability has not been studied, and a comparison of fully stable domains remains open.
We note only that at $\beta=2$ the radially stable black holes of the exponential completion, with
$M/\sqrt{\alpha}\geq1.068$, all lie in the range in which the quadratic truncation is stable at
least against $\ell=2$ and $\ell=4$ perturbations. Whether this persists for all multipoles, and
whether the exponential completion develops non-radial instabilities of its own, requires a
dedicated analysis.

Further limitations should be noted. Hyperbolicity has been examined for the principal part of the
radial master equation only, which does not establish the well-posedness of the full nonlinear
evolution. The coefficient $c_{4}(\beta)$ of the completion law has been computed numerically. Our
survey covers $0\leq\beta\leq5$ for a massless scalar, four slices of the massive and
self-interacting sector, four completions of the Ricci coupling, and a quadratic Gauss--Bonnet
coupling, and it is restricted to static, spherically symmetric black holes. Finally, the final
approach of the branches with $\gamma<0$ to the zero of $F$ has not been resolved numerically, and the
branches of the completion R have been followed beyond their turning points but not to their ends.
None of these limitations affects the conclusions stated below, but each marks a direction in
which they could be extended.

\subsection{Conclusions}
\label{subsec:conclusions}

We have studied how the nonlinear completion of the Ricci coupling shapes the spontaneous
scalarization of static black holes in scalar--Gauss--Bonnet gravity. Our main conclusions are as
follows.

First, the geometry of the horizon controls the existence of the branches. For arbitrary couplings
and potential, the near-horizon regularity condition and the horizon curvature invariants take
closed forms. Every branch of the completions Q and E terminates where the discriminant of the
regularity condition vanishes, at finite horizon curvature, and the regularity function reflects the
competition between the Einstein-frame kinetic term of the scalar and the Gauss--Bonnet coupling.

Second, the direction of the bifurcation, and with it the radial stability near the threshold, is
determined exactly. For a massless scalar, the tilt is a quadratic polynomial in $\beta$ that
changes sign at $\btilt=1.1156$, and a quartic self-interaction contributes a term linear in its
strength, with coefficient $c_{\lambda}=5.0021$. At second order, the Ricci coupling lowers the mass
of a black hole with given $r_{h}$ and $\phi_{h}$ by exactly $\beta\phi_{h}^{2}r_{h}/16$.

Third, at fourth order in the horizon value of the scalar the completions differ in mass by an
amount proportional to $\gamma$, with a coefficient that depends on $\beta$, and the regularity
bound shifts by an amount proportional to $-\gamma$. These two effects, a shift of the branch
toward larger masses and a shortening of the branch for $\gamma>0$, explain the ordering of the
domains of existence of different completions, and the predicted ratios $+2$ and $-1$ of the
coefficients for the rational and negative-$\gamma$ completions are confirmed to within
$3\times10^{-4}$.

Fourth, combined with the exchange of radial stability at the bifurcation and at turning points,
which we verify with a master equation for radial perturbations derived for arbitrary couplings,
these effects imply that the domain of radially stable scalarized black holes shrinks
monotonically as $\gamma$ increases over the range studied. In particular, the exponential
completion supports fewer radially stable black holes than the quadratic truncation, both without
and with a scalar mass and self-interaction, and the black holes by which the exponential
completion exceeds the quadratic truncation in its domain of existence are all radially unstable.

Fifth, these conclusions are independent of the conformal frame.

Several extensions suggest themselves: the non-radial stability of non-quadratic completions,
which would settle the comparison of fully stable domains; an analytic determination of
$c_{4}(\beta)$ from fourth-order perturbation theory; the resolution of the strongly coupled end of
branches with $\gamma<0$; the combination of completions of the Ricci and Gauss--Bonnet couplings;
and rotating and dynamically formed scalarized black holes, in which the role of $\gamma$ remains
to be explored.

\appendix

\section{Covariant field equations and their verification}
\label{app:covariant}
The field equations~\eqref{eq:Emn} and~\eqref{eq:Ephi} follow from the action~\eqref{eq:action}
by standard variation, with the Gauss--Bonnet contribution written in terms of the double
dual~\eqref{eq:P}. Because sign and normalization conventions for these equations differ across
the literature, we have verified them independently of their derivation.
\begin{enumerate}
\renewcommand{\labelenumi}{(\roman{enumi})}
\item \emph{Noether identity.} We evaluated both sides of Eq.~\eqref{eq:noether}, with the equations
exactly as written, on a static, non-vacuum test configuration with generic coupling functions:
$A=1-2/r+3/(10r^{2})$, $B=1-2/r+1/(5r^{3})$, $\phi=3/(10r)+1/(10r^{2})$,
$F=1-3\phi^{2}/10+\phi^{4}/5$, $f=2\phi^{2}/5$, and $V=\phi^{2}/10+\phi^{4}/20$, at
$r=37/10$ and $\theta=11/10$. Both sides equal $-2.5264589378\times10^{-4}$ for $\nu=r$ and agree
to 30 significant digits; the $\theta$ component vanishes identically. The identity also holds, to
$10^{-13}$, on random time-dependent, spherically symmetric configurations.
\item \emph{Double dual.} On the same configuration, $P_{\mu\rho\nu\sigma}$ has the algebraic
symmetries of the Riemann tensor, $P^{\mu}{}_{\rho\mu\sigma}=-G_{\rho\sigma}$, and
$\nabla^{\mu}P_{\mu\rho\nu\sigma}=0$, all exactly.
\item \emph{Reduced equations.} The reduced Lagrangian~\eqref{eq:L1D} differs from the
four-dimensional Lagrangian density evaluated on the ansatz~\eqref{eq:ansatz} by a total radial
derivative. Its Euler--Lagrange equations with respect to $\delta$, $N$, and $\phi$ are
proportional, with nonvanishing factors, to Eq.~\eqref{eq:N-eq}, Eq.~\eqref{eq:delta-eq}, and the
scalar equation, respectively. The expressions~\eqref{eq:GB-R-ansatz},
including the total-derivative form of $\GB$, coincide with the curvature scalars computed directly
from the metric.
\item \emph{Limits.} On the Schwarzschild background, $R=0$ and $\GB=48M^{2}/r^{6}$, and the
linearized scalar equation takes the form~\eqref{eq:linear}, independent of $\beta$ and $\gamma$.
\item \emph{Einstein frame.} The identity~\eqref{eq:einstein-frame} was verified by showing that
the Euler--Lagrange equations of its two sides coincide for a general static metric and arbitrary
$F$ and $V$; the expansions~\eqref{eq:KX} and the values of $\gamma$ in
Table~\ref{tab:completions} were verified symbolically.
\end{enumerate}

\section{Near-horizon expansions}
\label{app:horizon}
Substituting the expansion~\eqref{eq:frobenius} into Eqs.~\eqref{eq:delta-eq}, \eqref{eq:N-eq},
and the scalar equation, and expanding in $s=r-r_{h}$, the lowest order yields Eq.~\eqref{eq:b1}
for $b_{1}$. At the next order the three equations are linear in $(\delta_{1},b_{2},\phi_{2})$, but
their $3\times3$ coefficient matrix has identically vanishing determinant: the horizon is a regular
singular point of the system, because $N$ multiplies the highest radial derivatives. The system is
therefore solvable only if its right-hand side lies in the range of the matrix. After elimination of
$b_{1}$ by Eq.~\eqref{eq:b1} and removal of a nonvanishing factor, this condition becomes
\begin{equation}
A_{2}\phi_{1}^{2}+B_{2}\phi_{1}+C_{2}=0,\qquad
A_{2}=-16r_{h}\mathcal{P}_{h}\Pi_{h},\quad B_{2}=-32r_{h}^{2}\mathcal{B}_{h},\quad
C_{2}=64F_{h}r_{h}\mathcal{C}_{h},
\label{eq:quadratic-raw}
\end{equation}
which is Eq.~\eqref{eq:quadratic} multiplied by $-16r_{h}$. We have verified the factorized forms
of $A_{2}$, $B_{2}$, and $C_{2}$ and the polynomial identity
\begin{equation}
D_{2}\equiv B_{2}^{2}-4A_{2}C_{2}=1024\,r_{h}^{2}\bigl(2F_{h}-V_{h}r_{h}^{2}\bigr)^{2}\Delta_{h}
\label{eq:D2}
\end{equation}
symbolically, for arbitrary values of $F_{h}$, $F'_{h}$, $f'_{h}$, $V_{h}$, and $V'_{h}$, together
with $\mathcal{S}_{h}=r_{h}^{2}\mathcal{Q}_{h}-24\mathcal{P}_{h}f'_{h}$. For $V=0$,
$D_{2}=4096\,r_{h}^{2}F_{h}^{4}\mathcal{Q}_{h}\mathcal{S}_{h}$ and
$D_{2}/B_{2}^{2}=\mathcal{S}_{h}/(r_{h}^{2}\mathcal{Q}_{h})$, which is the normalized discriminant
used in Sec.~\ref{subsec:hyperbolicity}.

As $\phi_{h}\to0$ one has $C_{2}\to0$, and one root of Eq.~\eqref{eq:quadratic-raw} tends to zero
while the other remains finite. The physical root is the former; in rationalized form it is
Eq.~\eqref{eq:root}, which we have verified to solve the quadratic and to coincide with the root of
smallest modulus for random admissible data, with and without a potential, to
$6\times10^{-15}$. The rationalized form is regular at $\mathcal{P}_{h}=0$, where $A_{2}=0$ and the
second root diverges. Indeed, at $\mathcal{P}_{h}=0$ the identity above gives
$\mathcal{Q}_{h}=F_{h}r_{h}^{2}$, so that $\Delta_{h}=F_{h}^{4}r_{h}^{6}$ and
$\mathcal{B}_{h}=F_{h}^{2}r_{h}^{2}(2F_{h}-V_{h}r_{h}^{2})$. The denominator of
Eq.~\eqref{eq:root} then equals $2r_{h}\mathcal{B}_{h}$, and Eq.~\eqref{eq:root} reduces to
$\phi_{1}=2F_{h}\mathcal{C}_{h}/(r_{h}\mathcal{B}_{h})$, the solution of the linear equation to
which Eq.~\eqref{eq:quadratic} degenerates. The textbook expression
$(-B_{2}\pm\sqrt{D_{2}})/(2A_{2})$, with the root of smallest modulus selected, loses precision as
$\mathcal{P}_{h}\to0$. Away from this point the two forms agree to machine precision, and for the
branches of the completion M, which approach $\mathcal{P}_{h}=0$ (Sec.~\ref{sec:mechanism}), we use
the rationalized form.

We have verified the following reductions symbolically. For $F=1$ and $V=0$,
$\mathcal{Q}_{h}\mathcal{S}_{h}=r_{h}^{2}(r_{h}^{4}-96f_{h}'^{2})$. For quadratic Ricci and
Gauss--Bonnet couplings with $V=0$, the physical root coincides with that of
Ref.~\cite{Antoniou2021}. For $F=1$ with a potential, $\Delta_{h}=r_{h}^{6}\Delta_{\mathrm{M}}$, where
$\Delta_{\mathrm{M}}$ is the discriminant of Ref.~\cite{Macedo2019} expressed in the present
variables. The orthonormal components~\eqref{eq:riemann-h} follow from the expansion of the metric
functions to second order, and the closed form~\eqref{eq:X} follows from the scalar equation on
the horizon. Along the physical root the bracket in Eq.~\eqref{eq:X} vanishes at
$\mathcal{P}_{h}=0$, and $X$ is finite and continuous there, with and without a potential.

\section{Integrals of the weakly nonlinear expansion}
\label{app:integrals}
With the notation of Sec.~\ref{sec:wnl}, $F_{2}=-\beta\phi_{1}^{2}/4$, and
$f_{2}=\zeta_{0}\phi_{1}^{2}/2$, the second-order metric functions follow from
Eqs.~\eqref{eq:delta-eq} and~\eqref{eq:N2-eq}:
\begin{align}
\delta_{2}'&=\frac{r}{4}\,\phi_{1}'^{2}+\frac{r}{2}\,F_{2}''+\frac{2}{r^{2}}\,f_{2}'',
\label{eq:delta2}\\
rN_{2}&=-\frac{1}{4}\biggl\{\int_{1}^{r}\Bigl[\rho^{2}N_{0}\phi_{1}'^{2}-16N_{0}(N_{0}-1)f_{2}''
-8(3N_{0}-1)N_{0}'f_{2}'\Bigr]d\rho+4r^{2}N_{0}F_{2}'-2F_{2}-\frac{\beta}{2}\biggr\},
\label{eq:N2}
\end{align}
where the integrand is evaluated at $\rho$; Eq.~\eqref{eq:N2} satisfies $N_{2}(1)=0$. At third order the scalar
equation takes the form $\mathsf{L}[\phi_{3}]+S_{3}=0$, with
\begin{equation}
S_{3}=\bigl(r^{2}N_{2}\phi_{1}'\bigr)'+r^{2}N_{0}\delta_{2}'\phi_{1}'+\zeta_{0}r^{2}\phi_{1}\GB^{(2)}
+\frac{12\zeta_{2}}{r^{4}}\,\phi_{1}-\frac{\beta}{2}\,r^{2}\phi_{1}R^{(2)}-2\lambda r^{2}\phi_{1}^{3},
\label{eq:S3}
\end{equation}
where $R^{(2)}$ and $\GB^{(2)}$ are the second-order parts of Eq.~\eqref{eq:GB-R-ansatz} evaluated
with $N=N_{0}+\varepsilon^{2}N_{2}$ and $\delta=\varepsilon^{2}\delta_{2}$. We have verified
Eqs.~\eqref{eq:delta2} and~\eqref{eq:S3} symbolically by expanding the field equations. Inserting
Eq.~\eqref{eq:S3} into the solvability condition~\eqref{eq:fredholm} and integrating the first term by
parts, with boundary terms that vanish because $N_{2}(1)=0$ and $\phi_{1}\simeq Q_{1}/r$, gives
\begin{equation}
\zeta_{2}=\frac{1}{I_{z}}\int_{1}^{\infty}\Bigl[r^{2}N_{2}\phi_{1}'^{2}-r^{2}N_{0}\delta_{2}'\phi_{1}\phi_{1}'
-\zeta_{0}r^{2}\phi_{1}^{2}\GB^{(2)}+\frac{\beta}{2}\,r^{2}\phi_{1}^{2}R^{(2)}+2\lambda r^{2}\phi_{1}^{4}\Bigr]dr .
\label{eq:zeta2}
\end{equation}
Since $N_{2}$, $\delta_{2}$, $R^{(2)}$, and $\GB^{(2)}$ are at most linear in $\beta$, this expression
makes the quadratic dependence of $\zeta_{2}$ on $\beta$, and the linear dependence on $\lambda$ with
coefficient $2J_{4}/I_{z}$, explicit.

For a massless scalar without self-interaction,
$\zeta_{2}(\beta)=-0.0253413702-0.0265540280\,\beta+0.0563903253\,\beta^{2}$, and the integrals of
Sec.~\ref{sec:wnl} take the values $I_{z}=1.7441821390$, $J_{4}=0.2445619248$, and $0.0907016061$ for
the integral in Eq.~\eqref{eq:M2}. All integrals converge, since $\phi_{1}\simeq Q_{1}/r$ at large
$r$. We have evaluated them in two independent ways: by direct symbolic and numerical
implementation of the second- and third-order equations, and by quadrature of the entire
series~\eqref{eq:series}, truncated at 400 terms, where $|c_{400}|\sim10^{-529}$, in the variable
$v=1-1/r$ with 40-digit arithmetic. The two agree to nine or more significant digits.

\section{Numerical methods and error control}
\label{app:numerics}
\emph{Static solutions.} We integrate the field equations in the gauge
$ds^{2}=-A\,dt^{2}+dr^{2}/B+r^{2}d\Omega^{2}$, using $\ln A$, $B$, $\phi$, and $\phi'$ as variables,
from $r=r_{h}(1+\epsilon)$ with the horizon data of Sec.~\ref{subsec:regularity}, with an
eighth-order Runge--Kutta (DOP853) method at relative and absolute tolerances $10^{-11}$ and
$10^{-13}$, to $r_{\max}=2\times10^{4}r_{h}$. There, the asymptotic value of the scalar is extracted
from its asymptotic expansion including the $O(r^{-2})$ term, and $\zeta=\alpha/r_{h}^{2}$ is
determined by Brent's method from the condition that this value vanish. The dependence on the
horizon offset is removed by Richardson extrapolation over
$\epsilon\in\{4,2,1\}\times10^{-4}$. An independent second integration route reproduces $\zeta$ to
$2\times10^{-13}$. For $\mu>0$ we integrate to
$\mu r_{\mathrm{stop}}=20$, so that the decaying and growing Yukawa solutions differ by a factor
$e^{40}$, detect the growing solution explicitly, and read off the mass from the plateau of the mass
function before the growing solution sets in.

\emph{Branches and endpoints.} Branches are followed in $\phi_{h}$, with the bracket for $\zeta$
predicted by polynomial extrapolation along the branch. Near turning points and endpoints we use
pseudo-arclength continuation, with a predictor along the tangent, a Newton corrector with
central-difference derivatives, and turning points detected by a sign change of the
$\phi_{h}$ component of the tangent. The endpoint of a branch is located where the regularity
condition ceases to admit a real root, by bisection in $\phi_{h}$, followed by a dense resolution
where margins are small; for slice S1a at $\beta=0.3$, for example, this gives the margin
$(4.507\pm0.018)\times10^{-5}$ quoted in Sec.~\ref{subsec:massive-sector}.

\emph{Accuracy.} Besides the agreement of the two integration routes, the solutions satisfy the
following checks. The threshold~\eqref{eq:threshold} is reproduced to $1.7\times10^{-9}$ in
$\zeta$. The first law $dM=T\,dS$ with the Wald entropy~\eqref{eq:T-S} holds to relative accuracies
between $3\times10^{-9}$ and $2.4\times10^{-7}$, whereas it is clearly violated if the area law is
used instead of the Wald entropy. The residuals of the constraint are of order
$10^{-10}$. For $\beta=0$ the completions yield identical results. The edges of the domains of
existence are accurate to about $10^{-5}$ in $\hM$, and all margins quoted in
Sec.~\ref{sec:existence} exceed the corresponding uncertainties by about an order of magnitude or
more.

\emph{Perturbations.} For the stability analysis the background is recomputed with dense output
($\epsilon=10^{-5}$, $r_{\max}=2\times10^{4}r_{h}$), and the coefficients of
Eq.~\eqref{eq:master} are evaluated on it. The growth indicator measures the component of the
horizon-regular solution along the solution that grows at infinity; its zeros in $\Omega$ are the
unstable modes. The Sturm count uses the static solution that is regular on the horizon. Where very
slowly growing modes must be resolved, the radial domain is extended to $1.8\times10^{5}r_{h}$. The
normalized speed $s(r)$ is sampled on logarithmic grids near the horizon, starting at
$r-r_{h}=10^{-4}r_{h}$, and on linear grids farther out.

\section{Structure of the radial master equation}
\label{app:master}
The master equation~\eqref{eq:master} is obtained in three steps. First, the linearized $tr$
equation is proportional to $\partial_{t}X$, with
\begin{equation}
X=c_{1}b+c_{2}q+c_{3}q',
\label{eq:X-tr}
\end{equation}
where, with $F_{k}\equiv d^{k}F/d\phi^{k}$ and $f_{k}\equiv d^{k}f/d\phi^{k}$ evaluated at
$\phi_{0}(r)$,
\begin{align}
c_{1}&=-\frac{F_{0}}{r}-\frac{F_{1}\phi_{0}'}{2}+\frac{2f_{1}\phi_{0}'}{r^{2}}\,(3B_{0}-1),\qquad
c_{3}=-\frac{\mathcal{P}(r)}{r^{2}},\nonumber\\
c_{2}&=-\frac{\phi_{0}'}{2}-\phi_{0}'\Bigl[F_{2}+\frac{4f_{2}}{r^{2}}\,(1-B_{0})\Bigr]
+\frac{A_{0}'}{2A_{0}}\,\frac{\mathcal{P}(r)}{r^{2}},\qquad
\mathcal{P}(r)\equiv r^{2}F_{1}+4f_{1}(1-B_{0}).
\label{eq:c123}
\end{align}
For perturbations with nontrivial time dependence, $X=0$ determines $b$. The function
$\mathcal{P}(r)$ is the radial extension of the degeneracy factor of Sec.~\ref{subsec:regularity}:
on the horizon, where $B_{0}=0$, it reduces to $\mathcal{P}_{h}$, so that the vanishing of the
leading coefficient of the regularity condition coincides with the disappearance of the $q'$ term
from the momentum constraint. In general relativity with a minimally coupled scalar, $c_{1}=-1/r$,
$c_{2}=-\phi_{0}'/2$, and $c_{3}=0$, so that $b=-r\phi_{0}'q/2$. Second, the linearized $rr$
equation determines $a'$ algebraically in terms of $b$, $q$, $q'$, and $\partial_{t}^{2}q$. Third,
substituting $b$ and $a'$ into the linearized scalar equation yields Eq.~\eqref{eq:master}.

The coefficient of $\partial_{t}^{2}\partial_{r}q$ generated in the intermediate steps vanishes on
backgrounds that satisfy the static field equations, to $10^{-17}$ relative accuracy in our
numerical evaluations. The coefficients $K_{tt}$, $K_{rr}$, $K_{r}$, and $K_{0}$ are rational
functions of $A_{0}$, $A_{0}'$, $A_{0}''$, $B_{0}$, $B_{0}'$, $\phi_{0}'$, $\phi_{0}''$ and of $F$,
$f$, and $V$ and their derivatives up to third order, which we generate and evaluate by computer
algebra. Their explicit form is lengthy, since $K_{0}$ alone comprises more than $10^{4}$
operations, and we do not reproduce it here; the procedure above determines it uniquely. The limits
used to validate the coefficients are listed in Sec.~\ref{subsec:master}.



\begin{thebibliography}{44}

\bibitem{DEF1993}
T.~Damour and G.~Esposito-Far\`{e}se, \emph{Nonperturbative strong-field effects in tensor-scalar theories of gravitation}, \href{https://doi.org/10.1103/PhysRevLett.70.2220}{Phys.\ Rev.\ Lett. \textbf{70}, 2220 (1993)}.

\bibitem{DEF1996}
T.~Damour and G.~Esposito-Far\`{e}se, \emph{Tensor-scalar gravity and binary-pulsar experiments}, \href{https://doi.org/10.1103/PhysRevD.54.1474}{Phys.\ Rev.\ D \textbf{54}, 1474 (1996)}, \href{https://arxiv.org/abs/gr-qc/9602056}{arXiv:gr-qc/9602056}.

\bibitem{DonevaYazadjiev2018}
D.~D. Doneva and S.~S. Yazadjiev, \emph{New Gauss-Bonnet Black Holes with Curvature-Induced Scalarization in Extended Scalar-Tensor Theories}, \href{https://doi.org/10.1103/PhysRevLett.120.131103}{Phys.\ Rev.\ Lett. \textbf{120}, 131103 (2018)}, \href{https://arxiv.org/abs/1711.01187}{arXiv:1711.01187}.

\bibitem{Silva2018}
H.~O. Silva, J.~Sakstein, L.~Gualtieri, T.~P. Sotiriou, and E.~Berti, \emph{Spontaneous scalarization of black holes and compact stars from a Gauss-Bonnet coupling}, \href{https://doi.org/10.1103/PhysRevLett.120.131104}{Phys.\ Rev.\ Lett. \textbf{120}, 131104 (2018)}, \href{https://arxiv.org/abs/1711.02080}{arXiv:1711.02080}.

\bibitem{AntoniouBakopoulosKanti2018}
G.~Antoniou, A.~Bakopoulos, and P.~Kanti, \emph{Evasion of No-Hair Theorems and Novel Black-Hole Solutions in Gauss-Bonnet Theories}, \href{https://doi.org/10.1103/PhysRevLett.120.131102}{Phys.\ Rev.\ Lett. \textbf{120}, 131102 (2018)}, \href{https://arxiv.org/abs/1711.03390}{arXiv:1711.03390}.

\bibitem{Herdeiro2018charged}
C.~A.~R. Herdeiro, E.~Radu, N.~Sanchis-Gual, and J.~A. Font, \emph{Spontaneous Scalarization of Charged Black Holes}, \href{https://doi.org/10.1103/PhysRevLett.121.101102}{Phys.\ Rev.\ Lett. \textbf{121}, 101102 (2018)}, \href{https://arxiv.org/abs/1806.05190}{arXiv:1806.05190}.

\bibitem{Dima2020}
A.~Dima, E.~Barausse, N.~Franchini, and T.~P. Sotiriou, \emph{Spin-induced black hole spontaneous scalarization}, \href{https://doi.org/10.1103/PhysRevLett.125.231101}{Phys.\ Rev.\ Lett. \textbf{125}, 231101 (2020)}, \href{https://arxiv.org/abs/2006.03095}{arXiv:2006.03095}.

\bibitem{Doneva2024RMP}
D.~D. Doneva, F.~M. Ramazano\u{g}lu, H.~O. Silva, T.~P. Sotiriou, and S.~S. Yazadjiev, \emph{Spontaneous scalarization}, \href{https://doi.org/10.1103/RevModPhys.96.015004}{Rev.\ Mod.\ Phys. \textbf{96}, 015004 (2024)}, \href{https://arxiv.org/abs/2211.01766}{arXiv:2211.01766}.

\bibitem{Kanti1996}
P.~Kanti, N.~E. Mavromatos, J.~Rizos, K.~Tamvakis, and E.~Winstanley, \emph{Dilatonic black holes in higher curvature string gravity}, \href{https://doi.org/10.1103/PhysRevD.54.5049}{Phys.\ Rev.\ D \textbf{54}, 5049 (1996)}, \href{https://arxiv.org/abs/hep-th/9511071}{arXiv:hep-th/9511071}.

\bibitem{SotiriouZhou2014}
T.~P. Sotiriou and S.-Y. Zhou, \emph{Black hole hair in generalized scalar-tensor gravity}, \href{https://doi.org/10.1103/PhysRevLett.112.251102}{Phys.\ Rev.\ Lett. \textbf{112}, 251102 (2014)}, \href{https://arxiv.org/abs/1312.3622}{arXiv:1312.3622}.

\bibitem{BlazquezSalcedo2018}
J.~L. Bl\'{a}zquez-Salcedo, D.~D. Doneva, J.~Kunz, and S.~S. Yazadjiev, \emph{Radial perturbations of the scalarized Einstein-Gauss-Bonnet black holes}, \href{https://doi.org/10.1103/PhysRevD.98.084011}{Phys.\ Rev.\ D \textbf{98}, 084011 (2018)}, \href{https://arxiv.org/abs/1805.05755}{arXiv:1805.05755}.

\bibitem{Silva2019}
H.~O. Silva, C.~F.~B. Macedo, T.~P. Sotiriou, L.~Gualtieri, J.~Sakstein, and E.~Berti, \emph{Stability of scalarized black hole solutions in scalar-Gauss-Bonnet gravity}, \href{https://doi.org/10.1103/PhysRevD.99.064011}{Phys.\ Rev.\ D \textbf{99}, 064011 (2019)}, \href{https://arxiv.org/abs/1812.05590}{arXiv:1812.05590}.

\bibitem{Macedo2019}
C.~F.~B. Macedo, J.~Sakstein, E.~Berti, L.~Gualtieri, H.~O. Silva, and T.~P. Sotiriou, \emph{Self-interactions and spontaneous black hole scalarization}, \href{https://doi.org/10.1103/PhysRevD.99.104041}{Phys.\ Rev.\ D \textbf{99}, 104041 (2019)}, \href{https://arxiv.org/abs/1903.06784}{arXiv:1903.06784}.

\bibitem{Antoniou2021}
G.~Antoniou, A.~Leh\'{e}bel, G.~Ventagli, and T.~P. Sotiriou, \emph{Black hole scalarization with Gauss-Bonnet and Ricci scalar couplings}, \href{https://doi.org/10.1103/PhysRevD.104.044002}{Phys.\ Rev.\ D \textbf{104}, 044002 (2021)}, \href{https://arxiv.org/abs/2105.04479}{arXiv:2105.04479}.

\bibitem{Antoniou2022}
G.~Antoniou, C.~F.~B. Macedo, R.~McManus, and T.~P. Sotiriou, \emph{Stable spontaneously-scalarized black holes in generalized scalar-tensor theories}, \href{https://doi.org/10.1103/PhysRevD.106.024029}{Phys.\ Rev.\ D \textbf{106}, 024029 (2022)}, \href{https://arxiv.org/abs/2204.01684}{arXiv:2204.01684}.

\bibitem{Doneva2019massive}
D.~D. Doneva, K.~V. Staykov, and S.~S. Yazadjiev, \emph{Gauss-Bonnet black holes with a massive scalar field}, \href{https://doi.org/10.1103/PhysRevD.99.104045}{Phys.\ Rev.\ D \textbf{99}, 104045 (2019)}, \href{https://arxiv.org/abs/1903.08119}{arXiv:1903.08119}.

\bibitem{Andreou2019}
N.~Andreou, N.~Franchini, G.~Ventagli, and T.~P. Sotiriou, \emph{Spontaneous scalarization in generalized scalar-tensor theory}, \href{https://doi.org/10.1103/PhysRevD.99.124022}{Phys.\ Rev.\ D \textbf{99}, 124022 (2019)}; \textbf{101}, 109903(E) (2020), \href{https://arxiv.org/abs/1904.06365}{arXiv:1904.06365}.

\bibitem{Thaalba2024}
F.~Thaalba, M.~Bezares, N.~Franchini, and T.~P. Sotiriou, \emph{Spherical collapse in scalar-Gauss-Bonnet gravity: Taming ill-posedness with a Ricci coupling}, \href{https://doi.org/10.1103/PhysRevD.109.L041503}{Phys.\ Rev.\ D \textbf{109}, L041503 (2024)}, \href{https://arxiv.org/abs/2306.01695}{arXiv:2306.01695}.

\bibitem{Thaalba2025}
F.~Thaalba, N.~Franchini, M.~Bezares, and T.~P. Sotiriou, \emph{Dynamics of spherically symmetric black holes in scalar-Gauss-Bonnet gravity with a Ricci coupling}, \href{https://doi.org/10.1103/PhysRevD.111.064054}{Phys.\ Rev.\ D \textbf{111}, 064054 (2025)}, \href{https://arxiv.org/abs/2409.11398}{arXiv:2409.11398}.

\bibitem{RipleyPretorius2019a}
J.~L. Ripley and F.~Pretorius, \emph{Hyperbolicity in spherical gravitational collapse in a Horndeski theory}, \href{https://doi.org/10.1103/PhysRevD.99.084014}{Phys.\ Rev.\ D \textbf{99}, 084014 (2019)}, \href{https://arxiv.org/abs/1902.01468}{arXiv:1902.01468}.

\bibitem{RipleyPretorius2019b}
J.~L. Ripley and F.~Pretorius, \emph{Gravitational collapse in Einstein dilaton-Gauss--Bonnet gravity}, \href{https://doi.org/10.1088/1361-6382/ab2416}{Class.\ Quantum Grav.\ \textbf{36}, 134001 (2019)}, \href{https://arxiv.org/abs/1903.07543}{arXiv:1903.07543}.

\bibitem{EastRipley2021}
W.~E. East and J.~L. Ripley, \emph{Evolution of Einstein-scalar-Gauss-Bonnet gravity using a modified harmonic formulation}, \href{https://doi.org/10.1103/PhysRevD.103.044040}{Phys.\ Rev.\ D \textbf{103}, 044040 (2021)}, \href{https://arxiv.org/abs/2011.03547}{arXiv:2011.03547}.

\bibitem{Ventagli2021}
G.~Ventagli, G.~Antoniou, A.~Leh\'{e}bel, and T.~P. Sotiriou, \emph{Neutron star scalarization with Gauss-Bonnet and Ricci scalar couplings}, \href{https://doi.org/10.1103/PhysRevD.104.124078}{Phys.\ Rev.\ D \textbf{104}, 124078 (2021)}, \href{https://arxiv.org/abs/2111.03644}{arXiv:2111.03644}.

\bibitem{Herdeiro2026}
C.~Herdeiro, H.~Huang, J.~Kunz, M.-Y. Lai, E.~Radu, and D.-C. Zou, \emph{Phase structure of scalarized black holes in Einstein-scalar-Gauss-Bonnet gravity}, \href{https://doi.org/10.1103/4k1l-jzvz}{Phys.\ Rev.\ D \textbf{113}, 124046 (2026)}, \href{https://arxiv.org/abs/2603.24164}{arXiv:2603.24164}.

\bibitem{Staykov2025}
K.~V. Staykov, D.~D. Doneva, P.~G.~S. Fernandes, and S.~S. Yazadjiev, \emph{Rotating scalarized black holes: The role of the coupling}, \href{https://doi.org/10.1103/z8lr-4tb5}{Phys.\ Rev.\ D \textbf{112}, 024058 (2025)}, \href{https://arxiv.org/abs/2503.15206}{arXiv:2503.15206}.

\bibitem{Kleihaus2023}
B.~Kleihaus, J.~Kunz, T.~Uterm\"{o}hlen, and E.~Berti, \emph{Quadrupole instability of static scalarized black holes}, \href{https://doi.org/10.1103/PhysRevD.107.L081501}{Phys.\ Rev.\ D \textbf{107}, L081501 (2023)}, \href{https://arxiv.org/abs/2303.04107}{arXiv:2303.04107}.

\bibitem{Minamitsuji2024}
M.~Minamitsuji, S.~Mukohyama, and S.~Tsujikawa, \emph{Angular and radial stabilities of spontaneously scalarized black holes in the presence of scalar-Gauss-Bonnet couplings}, \href{https://doi.org/10.1103/PhysRevD.109.104057}{Phys.\ Rev.\ D \textbf{109}, 104057 (2024)}, \href{https://arxiv.org/abs/2403.10048}{arXiv:2403.10048}.

\bibitem{Rossi2026}
D.~Rossi, L.~Gualtieri, and T.~P. Sotiriou, \emph{Minimum mass, maximum charge, and hyperbolicity in scalar Gauss-Bonnet gravity}, \href{https://doi.org/10.1103/2qq5-cqzx}{Phys.\ Rev.\ D \textbf{114}, 064042 (2026)}, \href{https://arxiv.org/abs/2604.08668}{arXiv:2604.08668}.

\bibitem{Horndeski1974}
G.~W. Horndeski, \emph{Second-order scalar-tensor field equations in a four-dimensional space}, \href{https://doi.org/10.1007/BF01807638}{Int.\ J.\ Theor.\ Phys.\ \textbf{10}, 363 (1974)}.

\bibitem{Kobayashi2011}
T.~Kobayashi, M.~Yamaguchi, and J.~Yokoyama, \emph{Generalized G-inflation}, \href{https://doi.org/10.1143/PTP.126.511}{Prog.\ Theor.\ Phys.\ \textbf{126}, 511 (2011)}, \href{https://arxiv.org/abs/1105.5723}{arXiv:1105.5723}.

\bibitem{CrandallRabinowitz1971}
M.~G. Crandall and P.~H. Rabinowitz, \emph{Bifurcation from simple eigenvalues}, \href{https://doi.org/10.1016/0022-1236(71)90015-2}{J.\ Funct.\ Anal.\ \textbf{8}, 321 (1971)}.

\bibitem{Muniz2025}
J.~V.~M. Muniz, N.~Ortiz, and R.~F.~P. Mendes, \emph{Phase transition mechanism of spontaneous scalarization}, \href{https://doi.org/10.1103/qh47-qv76}{Phys.\ Rev.\ D \textbf{112}, 064037 (2025)}, \href{https://arxiv.org/abs/2503.11385}{arXiv:2503.11385}.

\bibitem{Unluturk2025}
K.~\.{I}. \"{U}nl\"{u}t\"{u}rk, S.~Tuna, O.~K. Yamak, and F.~M. Ramazano\u{g}lu, \emph{Nature of Phase Transitions and Metastability in Scalar-Tensor Theories}, \href{https://doi.org/10.1103/14xz-d7xf}{Phys.\ Rev.\ Lett. \textbf{135}, 061401 (2025)}, \href{https://arxiv.org/abs/2502.01781}{arXiv:2502.01781}.

\bibitem{Ozinan2026}
M.~\"{O}zinan, K.~\.{I}. \"{U}nl\"{u}t\"{u}rk, and F.~M. Ramazano\u{g}lu, \emph{Underlying mechanisms of phase transitions in scalar-tensor theories}, \href{https://doi.org/10.1103/1brh-qzls}{Phys.\ Rev.\ D \textbf{114}, 024054 (2026)}, \href{https://arxiv.org/abs/2604.24867}{arXiv:2604.24867}.

\bibitem{Sorkin1981}
R.~Sorkin, \emph{A criterion for the onset of instability at a turning point}, Astrophys.\ J.\ \textbf{249}, 254 (1981).

\bibitem{BlazquezSalcedo2024GRG}
J.~L. Bl\'{a}zquez-Salcedo, B.~Kleihaus, and J.~Kunz, \emph{Instabilities of black holes in Einstein-scalar--Gauss--Bonnet theories}, \href{https://doi.org/10.1007/s10714-024-03278-w}{Gen.\ Relativ.\ Gravit.\ \textbf{56}, 99 (2024)}.

\bibitem{BlazquezSalcedo2026hier}
J.~L. Bl\'{a}zquez-Salcedo, L.~M. Gonz\'{a}lez-Romero, F.~S. Khoo, J.~Kunz, and P.~Navarro Moreno, \emph{Hierarchy of Angular Instabilities in Scalarized Black Holes}, \href{https://arxiv.org/abs/2607.07208}{arXiv:2607.07208}.

\bibitem{JacobsonKang1993}
T.~Jacobson and G.~Kang, \emph{Conformal invariance of black hole temperature}, \href{https://doi.org/10.1088/0264-9381/10/11/002}{Class.\ Quantum Grav.\ \textbf{10}, L201 (1993)}, \href{https://arxiv.org/abs/gr-qc/9307002}{arXiv:gr-qc/9307002}.

\bibitem{Wald1993}
R.~M. Wald, \emph{Black hole entropy is the Noether charge}, \href{https://doi.org/10.1103/PhysRevD.48.R3427}{Phys.\ Rev.\ D \textbf{48}, R3427 (1993)}, \href{https://arxiv.org/abs/gr-qc/9307038}{arXiv:gr-qc/9307038}.

\bibitem{IyerWald1994}
V.~Iyer and R.~M. Wald, \emph{Some properties of the Noether charge and a proposal for dynamical black hole entropy}, \href{https://doi.org/10.1103/PhysRevD.50.846}{Phys.\ Rev.\ D \textbf{50}, 846 (1994)}, \href{https://arxiv.org/abs/gr-qc/9403028}{arXiv:gr-qc/9403028}.

\bibitem{MyungZou2019}
Y.~S. Myung and D.-C. Zou, \emph{Stability of scalarized charged black holes in the Einstein--Maxwell--Scalar theory}, \href{https://doi.org/10.1140/epjc/s10052-019-7176-7}{Eur.\ Phys.\ J.\ C \textbf{79}, 641 (2019)}, \href{https://arxiv.org/abs/1904.09864}{arXiv:1904.09864}.

\bibitem{KMS2012}
T.~Kobayashi, H.~Motohashi, and T.~Suyama, \emph{Black hole perturbation in the most general scalar-tensor theory with second-order field equations: The odd-parity sector}, \href{https://doi.org/10.1103/PhysRevD.85.084025}{Phys.\ Rev.\ D \textbf{85}, 084025 (2012)}; \textbf{96}, 109903(E) (2017), \href{https://arxiv.org/abs/1202.4893}{arXiv:1202.4893}.

\bibitem{KMS2014}
T.~Kobayashi, H.~Motohashi, and T.~Suyama, \emph{Black hole perturbation in the most general scalar-tensor theory with second-order field equations. II. The even-parity sector}, \href{https://doi.org/10.1103/PhysRevD.89.084042}{Phys.\ Rev.\ D \textbf{89}, 084042 (2014)}, \href{https://arxiv.org/abs/1402.6740}{arXiv:1402.6740}.

\bibitem{BlazquezSalcedo2026rev}
J.~L. Bl\'{a}zquez-Salcedo, F.~S. Khoo, B.~Kleihaus, and J.~Kunz, \emph{Black holes in alternative theories of gravity}, \href{https://arxiv.org/abs/2609.03983}{arXiv:2609.03983}.

\end{thebibliography}
\end{document}